\documentclass[preprintnumbers, floatfix, letterpaper, twocolumn,aps,prd, nofootinbib]{revtex4-1}
\pdfoutput=1
\usepackage{bm,graphicx,dcolumn,epstopdf,epsf, latexsym,mathbbol, amssymb,amsmath,color,slashed, mathrsfs,mathcomp,simplewick}
\usepackage[center]{subfigure}
\usepackage{multirow}
\usepackage{makecell}
\usepackage{float}
\usepackage{booktabs}
\usepackage[colorlinks,linkcolor=blue,citecolor=blue,urlcolor=blue]{hyperref}
\begin{document}
 \newcommand{\bq}{\begin{equation}} 
 \newcommand{\eq}{\end{equation}}
 \newcommand{\bqn}{\begin{eqnarray}}
 \newcommand{\eqn}{\end{eqnarray}}
 \newcommand{\nb}{\nonumber}
 \newcommand{\lb}{\label}
 \newcommand{\f}{\frac}
 \newcommand{\p}{\partial}
\newcommand{\PRL}{Phys. Rev. Lett.}
\newcommand{\PLB}{Phys. Lett. B}
\newcommand{\PRD}{Phys. Rev. D}
\newcommand{\CQG}{Class. Quantum Grav.}
\newcommand{\JCAP}{J. Cosmol. Astropart. Phys.}
\newcommand{\JHEP}{J. High. Energy. Phys.}
\newcommand{\Doi}{https://doi.org}
\newcommand{\arXiv}{https://arxiv.org/abs}
\newcommand{\red}{\textcolor{red}}
\title{Inflationary perturbation spectrum in extended effective field theory of inflation}

\author{Jin Qiao${}^{a}$}

\author{Guang-Hua Ding${}^{a}$}

\author{Qiang Wu${}^{a}$}

\author{Tao Zhu${}^{a}$}
\email{zhut05@zjut.edu.cn; Corresponding author}

\author{Anzhong Wang${}^{b}$}

\affiliation{${}^{a}$ Institute for Theoretical Physics $\&$ Cosmology, Zhejiang University of Technology, Hangzhou, 310032, China \\
${}^{b}$ GCAP-CASPER, Physics Department, Baylor University, Waco, TX 76798-7316, USA
}

\date{\today}

\begin{abstract}

The effective field theory (EFT) of inflation provides a natural framework to study the new physical effects on primordial perturbations. Recently a healthy extension of the EFT of inflation with high-order operators has been proposed, which avoids ghosts and meanwhile leads to a nonlinear dispersion relation of the scalar perturbations. This paper is devoted to studying the effects of these high-order operators by using the uniform asymptotic approximation method. In particular, we first construct the approximate analytical solution to the mode function of the scalar perturbations. Because of the presence of the high-order operators, the perturbation modes usually experience a period of non-adiabatic evolution before they cross the Hubble radius, which could lead to the production of excited states and modifications of the primordial perturbation spectrum. However, we show that the modified power spectrum is still nearly scale-invariant and the presence of the high-order operators can only affect the overall amplitude of the spectrum. In particular, after showing explicitly the impact of these new effects on particle production rate and perturbation spectrum, we explore their origin in detail.

\end{abstract}
\maketitle

\section{Introduction}
\renewcommand{\theequation}{1.\arabic{equation}}\setcounter{equation}{0}

The cosmic inflation has achieved remarkable successes not only in solving several fundamental and conceptual problems (such as the flatness,  horizon problem, and exotic relics) of the standard big bang cosmology, but more importantly, it provides a causal mechanism for generating the large-scale structure of the Universe and the cosmic microwave background (CMB) \cite{guth_inflationary_1981, starobinsky_new_1980, sato_firstorder_1981} (see Ref.~\cite{baumann_tasi_2009} for an updated review). All these predictions are  matched well to cosmological observations with a spectacular precision \cite{komatsu_sevenyear_2011, planckcollaboration_planck_2018, planck_collaboration_planck_2015-4, planck_collaboration_planck_2014-1}. These observations have provided a strong evidence of a nearly scale invariant power spectrum of adiabatic  perturbations and support the inflationary paradigm with a single scalar field, which gives rise to a slow-roll inflationary phase. During this slow-roll phase, the energy density of the matter field remains nearly constant and the spacetime behaves like a quasi-de Sitter spacetime. 

While there are a lot of approaches to realize the inflation with a single scalar field, the EFT of inflation provides a general framework for describing the most generic single scalar field theory and the associated fluctuations in a quasi de-Sitter background \cite{cheung_effective_2008, weinberg_effective_2008}. In this framework, the scalar field provides a clock that breaks time {diffeomorphism invariance but preserves the spatial one}. This allows one to construct the action of the theory around the quasi de-Sitter background in terms of  spatial diffeomorphism invariants and study the effects of different terms. This is very similar to the case of Ho\v{r}ava-Lifshitz (HL) theory of quantum gravity \cite{horava_quantum_2009, horava_general_2010, zhu_symmetry_2011, zhu_general_2012, lin_postnewtonian_2014}, in which the symmetry of the theory is broken from the general covariance down to the foliation-preserving diffeomorphisms. With this property, the action of the HL theory of quantum gravity has to be constructed in terms of 3-dimensional spatial diffeomorphism invariants. By including high-order spatial derivative operators (up to the six order), but excluding high-order temporal derivative operators, the HL theory of gravity becomes power-counting renormalizable. Such remarkable features have attracted a lot of attention and the observational effects of high-order operators {to} inflationary perturbation spectra have been extensively studied \cite{huang_primordial_2013, zhu_inflation_2013, zhu_effects_2013, wang_polarizing_2012} (see \cite{wang_horava_2017} for an updated review).

Considering the similarity between the two theories based on the same symmetry, it is natural to study these high-order  derivative terms  in the framework of EFT of inflation. These new terms provide an efficient way for parametrizing unknown high energy physics effects on the low energy {scale, and produce various} inflationary models \cite{tong_effective_2017, Gwyn:2012mw, Castillo:2013sfa, Hetz:2016ics, Gwyn:2014doa, alishahiha_dbi_2004, shandera_observing_2006, gong_higher_2015, arkani-hamed_ghost_2004}, such as inflation models in HL theory mentioned above, DBI inflation \cite{alishahiha_dbi_2004, shandera_observing_2006}, and Ghost inflation \cite{arkani-hamed_ghost_2004} (for EFT of bouncing universe, see \cite{cai_effective_2017, cai_higher_2017}). Recently, the EFT of inflation has been extended by adding higher spatial derivative terms (up to the fourth-order) \cite{ashoorioon_extended_2018}. With this extension, the usual linear dispersion relation associated with the propagation of inflationary scalar perturbation has been changed to a nonlinear one. The presence of the high-order operators also sets a new characteristic energy scale in the extended theory. Above this new energy scale, the high-order operators dominate, while below it the usual linear ones become dominant.

An important question now is whether the new high-order derivative operators in the extended EFT of inflation can leave any observational effects.  An essential step to address this issue is to investigate  the cosmological perturbations in the extended EFT of inflation and calculate the corresponding inflationary observables by evolving perturbation modes starting from the high energy regime where the higher-order terms dominate until the end of the slow-roll inflation. Such considerations have attracted a lot of attention recently \cite{ashoorioon_extended_2018, ashoorioon_getting_2017, ashoorioon_nonunitary_2018}. In particular, the impact of the resulting nonlinear dispersion relation on the primordial perturbation spectrum has been studied extensively (see \cite{ashoorioon_effects_2011, ashoorioon_note_2011, zhu_inflationary_2014, zhu_gravitational_2014-1, wu_CTP, zhu_quantum_2014, zhu_highorder_2016} and references therein). {One of the} effects is that the cosmological perturbations can experience a period of non-adiabatic evolution in the high energy regime, and {as a result},  the perturbations are no longer in the adiabatic Bunch-Davies state. These excited states in turn lead to particle production during inflation and modify the 
primordial perturbation spectrum.

According to the quantum field theory in curved spacetime, particle production can arise from non-adiabatic evolutions of the associated field modes \cite{winitzki_cosmological_2005}. {Indeed, this is exactly the case that occurs in} cosmological perturbations in the extended EFT of inflation \cite{ashoorioon_extended_2018, ashoorioon_getting_2017, ashoorioon_nonunitary_2018}. Such non-adiabatic {evolutions of} the primordial perturbations also {occur during the super-inflationary phase right after the quantum bounce in loop quantum cosmology, see,  for examples, \cite{wu_nonadiabatic_2018, zhu_primordial_2018a, zhu_preinflationary_2017, zhu_universal_2017b}. }However, when the adiabatic condition is violated, it is in general impossible to study exact solution and the corresponding power spectra for cosmological perturbations, and thus one has to use some approximate methods. Recently, we have developed a method, {\em the uniform asymptotic approximation method} \cite{zhu_constructing_2014, zhu_inflationary_2014, zhu_quantum_2014}, to calculate precisely the quantum gravitational effects of the primordial perturbation spectra. The robustness of this method has been verified for calculating primordial spectra in k-inflation \cite{zhu_power_2014, Wu:2017joj, martin_kinflationary_2013,ringeval_diracborninfeld_2010},  inflation with nonlinear dispersion relations \cite{zhu_inflationary_2014, zhu_quantum_2014, zhu_highorder_2016},  quantum gravitational effects in loop quantum cosmology \cite{zhu_scalar_2015, zhu_detecting_2015, zhu_inflationary_2016}, the parametric resonance during inflation and reheating \cite{Zhu:2018smk}, and applications to quantum mechanics \cite{Zhu:2019bwj}. We note here that this method was first applied to inflationary cosmology in the framework of GR in \cite{habib_inflationary_2002, habib_inflationary_2005, habib_characterizing_2004}, and then we {made various extensions}, so it can be applied to {other theories of gravity}, including the ones with nonlinear dispersion relations \cite{zhu_inflationary_2014,zhu_highorder_2016}. The main purpose of the present paper is to use this  method to derive the inflationary observables  of slow-roll inflation in the framework of the extended EFT. By analytically solving the equation of motion for scalar perturbations, we concentrate on how the high-order operators affect the evolution of the perturbation modes, by providing  the general expressions of theperturbation spectrum at the end of the slow-roll inflation. The main properties of the high-order operators have been discussed in detail. These expressions represent a significant improvement over the previous results obtained so far in the literature.

We organize the rest of the paper as follows. In Sec.~II, we provide a brief introduction to the extended EFT of inflation and the equations of motion for both cosmological scalar and tensor perturbations. In Sec.~III, we first discuss how the adiabatic condition is violated due to the introduction of  the high-order operators in the extended EFT of inflation, and then construct analytical solutions for scalar perturbation modes, by using the uniform asymptotic approximation method.  With the analytical solution we derive the general expression of the corresponding perturbation spectrum. In Sec.~IV,  we study the main features of both the particle production rate during the inflation and the primordial perturbation spectrum. Our main conclusions and discussions are presented in Sec.~V. The strong coupling issue in the extended EFT is also discussed in Appendix A.

\section{Extended effective field theory of inflation}
\renewcommand{\theequation}{2.\arabic{equation}}\setcounter{equation}{0}

In this section, we present a brief review of the extended EFT of inflation by including high-order derivative terms  \cite{ashoorioon_extended_2018}. In general, the EFT of inflation provides a framework for describing the most general single scalar field on a quasi de-Sitter background \cite{cheung_effective_2008, weinberg_effective_2008}. Since the Friedmann-Robterson-Walker (FRW) background metric provides a preferred time foliation, we can in general write the action of a theory around the FRW background in terms of only the 3-dimensional spatial diffeomorphism invariants.  Then,  it can be shown that the basic building blocks of this construction include scalars like $g^{00}$,  pure function of time $c(t)$, and the extrinsic curvature tensor $K^{\mu\nu}$ of the constant time hypersurfaces. Using these blocks, one can show that the action of the EFT of inflation around a flat FRW background reads
\begin{widetext}
\bqn\lb{Seff}
S_{\rm eft}&=&M_{\rm Pl}^2 \int d^4 x \sqrt{-g} \Bigg\{\frac{R}{2}+\dot H g^{00} - (3H^2+\dot H) + \frac{M_2^4}{2 M_{\rm Pl}^2}(g^{00}+1)^2 + \frac{\bar M_1^3}{2 M_{\rm Pl}^2} (g^{00}+1)\delta K^\mu_\mu   \nb\\
&& ~~~~~~~~~~~~~~~~~~~~~~  - \frac{\bar M_2^2}{2 M_{\rm Pl}^2} (\delta K_\mu^\mu)^2- \frac{\bar M_3^2}{2M_{\rm Pl}^2} \delta K^\mu_\nu \delta K^\nu_\mu\Bigg\},
\eqn
\end{widetext}
where $\delta K_{\mu\nu}$ denotes the perturbation of $K_{\mu\nu}$ about the flat FRW background.

The above action can be extended by including high-order derivative terms. In this paper, we follow the extension proposed in \cite{ashoorioon_extended_2018}, which includes only operators with spatial derivatives up to the fourth-order. The action for these additional terms is given by
\begin{widetext}
\bqn\lb{DeltaS}
\Delta S &=& \int d^4 x \sqrt{-g}\Bigg\{ \frac{\bar M_4}{2}\nabla g^{00} \nabla^\nu \delta K_{\mu\nu} -\frac{\delta_1}{2} (\nabla_{\mu} \delta K^{\nu\gamma})(\nabla^{\mu} \delta K_{\nu\gamma}) - \frac{\delta_2}{2}(\nabla_{\mu} \delta K^{\nu}_{\nu})^2 \nb\\
&&~~~~~~~~ ~~~~~~~~ - \frac{\delta_3}{2}(\nabla_{\mu} \delta K^{\mu}_{\nu})(\nabla_{\gamma} \delta K^{\gamma \nu}) - \frac{\delta_4}{2} \nabla^\mu \delta K_{\nu\mu} \nabla^\nu \delta K^{\sigma}_{\sigma}\Bigg\},
\eqn
\end{widetext}
where $\bar M_4$ {and} $\delta_i \; (i=1,2,3,4)$ are {constants}. The first term (the $\bar M_4$ term) contains three {derivatives}, thus it breaks the time reverse symmetry. The other terms ($\delta_i $ terms) contain {the fourth-order derivatives} and lead to sixth-order corrections to the standard linear dispersion relation of the scalar perturbations.

We would like to mention that, in writing the above action, one requires the perturbation in the inflaton field $\phi$ to be zero, i.e., $\delta \phi=0$. This is achieved by considering the following linear transformation,
\bqn
\tilde t = t+\xi^0(t,x^i), \;\;  \delta \tilde \phi = \delta  \phi + \xi^0(t, x^i) \dot \phi_0(t),
\eqn
which leads to a particular gauge (unitary gauge) with $\xi^0(t,x^i)= - \delta\phi/\dot \phi_0$, {in which}  there is no inflaton {perturbations. Obviously,  the actions of} equations (\ref{Seff}) and (\ref{DeltaS}), which  respect the unitary gauge, are not of  time diffeomorphism invariance. In order to restore such a symmetry, one can introduce a Goldstone mode $\pi(t,x^i)$ and require it to transform as $\pi(t,x^i) \to \pi(t,x^i)- \xi^0(t,x^i)$. Consequently, perturbation of the inflaton field is not required to be zero and it is related to $\pi(t,x^i)$ via $\delta \phi = \dot \phi_0 \pi$.

On the other hand, one must be careful when dealing with higher derivative operators in the theory. The main concern is that they in general produce time derivatives higher than second-order in the equations of motion, and according to Ostrogradski' s theorem, a system with higher order time derivatives is usually not free of ghosts. For this reason, as analyzed in \cite{ashoorioon_extended_2018}, one has to impose the condition
\bqn
\delta_1=0=\delta_2,
\eqn
in order to avoid the presence of high-order time derivatives. In this paper, we will adopt this condition and disregard the $\delta_1$ and $\delta_2$ terms in the above action.

\subsection{Tensor perturbations}

For the tensor perturbations, the perturbed spacetime is set to
\bqn
g_{ij}=a^2 (\delta_{ij}+h_{ij}),
\eqn
where $h_{ij}$ represents the transverse and traceless tensor perturbations, {and satisfies the conditions}, 
\bqn
h^{i}_{i}=0=\partial^i h_{ij}.
\eqn
Then,  expanding the total action $S_{\rm tot}=S_{\rm eft}+ \Delta S$ up to the second-order of {$h_{ij}$, we find }
\bqn
S^{2}_{h} = \frac{M_{\rm Pl}^2}{8} \int dt d^3x a^3 \left(c_t^{-2}\partial_t h_{ij} \partial_t h^{ij}-a^{-2} \partial_k h_{ij} \partial^k h^{ij}\right),\nb\\
\eqn
where the effective sound speed $c_t$ for the tensor perturbations is given by
\bqn
c_t^2 = \left(1-\frac{\bar M_3^2}{M_{\rm Pl}^2}\right)^{-1}.
\eqn
In order to avoid superluminal propagations for the tensor modes, we require
\bqn
\bar M_3 <0.
\eqn

Then, the variation of the total action with respect to $h_{ij}$ leads to the equation of motion
\bqn
\frac{d^2\mu_k^{(t)}(\eta)}{d\eta^2} + \left(c_t^2 k^2- \frac{a''}{a}\right) \mu_k^{(t)}(\eta)=0,
\eqn
where $d\eta =dt/a$ is the conformal time and $\mu_k^{(t)}(\eta)\equiv a M_{\rm Pl} h_k/\sqrt{2}$ with $h_k$ denoting the Fourier modes of  the tensor perturbations. In the slow-roll inflation, if we treat all the slow-rolling quantities approximately as constant, then the equation of motion for the tensor perturbations can be solved analytically and expressed as a linear combination of the Hankel functions,
\bqn
\mu_k^{(t)}(\eta) \simeq \frac{\sqrt{-\pi \eta}}{2} \left[\alpha_k H^{(1)}_{\nu_t} (-c_t k\eta) + \beta_k H^{(2)}_{\nu_t} (-c_t k\eta)\right],\nb\\
\eqn
where $H_{\nu_t}^{(1)}(-c_t k \eta)$ and $H_{\nu_t}^{(2)}(-c_t k \eta)$ denote the Hankel functions of the first and second kind, respectively, and $\nu_t$ is a slow-roll quantity defined as
\bqn
\nu_t^2 \equiv \eta^2 \frac{a''}{a}+\frac{1}{4}.
\eqn
The coefficients $\alpha_k$ and $\beta_k$ are two integration constants. In order to fix them, we impose the Bunch-Davis (BD) vacuum state at the initial time,
\bqn
\mu_k^{(t)}(\eta) = \frac{1}{\sqrt{2 kc_{t}}} e^{- ic_t k \eta},
\eqn
which leads to
\bqn
\alpha_k =1,\;\; \beta_k=0.
\eqn
Then,  the tensor perturbation spectrum is calculated at the end of inflation, i.e., $\eta \to 0^-$. By using the asymptotic form of the Hankel function when $\eta\to 0^-$, we obtain
\bqn
\mathcal{P}_h &\equiv& \frac{k^3}{\pi^2 M_{\rm Pl}^2}\left|\frac{\mu_k^{(t)}(\eta)}{a(\eta)}\right|^2 \nb\\
&=&\frac{k^2}{4c_t\pi^3}\frac{1}{a^2(\eta)}\Gamma^2(\nu_t) \left(-\frac{c_tk\eta}{2}\right)^{1-2\nu_t} .\nb\\
\eqn

\subsection{Scalar perturbations}

As mentioned above, the Goldstone mode $\pi(t,x^i)$ can be introduced to restore the time diffeomorphism invariance of the action. This also describes the scalar perturbations around the flat FRW background. Thus, in order to study the scalar perturbations, one can transform the action in (\ref{Seff}) and (\ref{DeltaS}) from the unitary gauge to the $\pi$-gauge by evaluating the action explicitly for $\pi$. Following this procedure, we find,
\bqn\lb{quadra_eff}
S^\pi &=&S_{\rm eff}^\pi+\Delta S^\pi = \int d^4x \sqrt{-g} (\mathcal{L}_{\rm eff}+\mathcal{L}_{\Delta S}),
\eqn
where
\bqn
\mathcal{L}_{\rm eff} &=& M^{2}_{\rm Pl}\dot{H}(\partial_{\mu}\pi)^{2}+2M^{4}_{2}\dot{\pi}^{2}-\bar{M}^{3}_{1}H \left(3\dot{\pi}^{2}-\frac{(\partial_{i}\pi)^{2}}{2a^{2}}\right)  \nb\\
&&-\frac{\bar{M}^{2}_{2}}{2}\left(9H^{2}\dot{\pi}^{2}-3H^{2}\frac{(\partial_{i}\pi)^{2}}{a^{2}}+\frac{(\partial^{2}_{i}\pi)^{2}}{a^{4}}\right)  \nb\\
&&-\frac{\bar{M}^{2}_{3}}{2} \left(3H^{2}\dot{\pi}^{2}-H^{2}\frac{(\partial_{i}\pi)^{2}}{a^{2}}+\frac{(\partial^{2}_{j}\pi)^{2}}{a^{4}}\right),
\eqn
and
\bqn
\mathcal{L}_{\Delta S} &=& \frac{\bar{M}_{4}}{2}\left(\frac{k^{4}H\pi^{2}}{a^{4}}+\frac{k^{2}H^{3}\pi^{2}}{a^{2}}-9H^{3}\dot{\pi}^{2}\right)   \nb\\
&&-\frac{1}{2}\delta_{3}\left(\frac{k^{6}\pi^{2}}{a^{6}}+\frac{3H^{2}k^{4}\pi^{2}}{a^{4}}+\frac{H^{2}k^{2}\dot{\pi}^{2}}{a^{2}}-9H^{4}\dot{\pi}^{2}\right)  \nb\\
&&-\frac{1}{2}\delta_{4} \Bigg(\frac{k^{6}\pi^{2}}{a^{6}}+\frac{H^{2}k^{4}\pi^{2}}{2a^{4}}+\frac{9H^{4}k^{2}\pi^{2}}{2a^{2}}+\frac{3H^{2}k^{2}\dot{\pi}^{2}}{a^{2}}  \nb\\
&&~~~~~~~~~ +\frac{27}{2}H^{4}\dot{\pi}^{2}\Bigg).
\eqn
Note that in the above equations only the quadratic terms in the action are included, since these are the most relevant parts for our discussions about the primordial power spectrum.

Variation of the above action $S^\pi$ with respect to $\pi$ yields the equation of motion for the Goldstone mode $\pi$,
\bqn
A_1 \ddot \pi_k + B_1 \dot \pi_k +\left(F_1 \frac{k^2}{a^2}+D_1 \frac{k^4}{a^4}+C_1 \frac{k^6}{a^6}\right)\pi_k=0,\nb\\
\eqn
where
\bqn
A_1 &=& -2M^{2}_{Pl}\dot{H}+4M^{4}_{2}-6\bar{M}^{3}_{1}H-9H^{2}\bar{M}^{2}_{2} \nb\\
&&-3H^{2}\bar{M}^{2}_{3}+2H^{4}F_{0}(k,\tau), \lb{A1}\\
B_1 &=&-6\bar{M}^{3}_{1}\dot{H}-18\dot{H}H\bar{M}^{2}_{2}-6\dot{H}H\bar{M}^{2}_{3} +3HC_{0}\nb\\
&&+H^{5}\left[6F_{0}(k,\tau)+\frac{2k^{2}}{a^{2}H^{2}}(\delta_{3}+3\delta_{4})\right], \\
C_1 &=&\delta_{3}+\delta_{4},  \\
D_1 &=&\bar{M}^{2}_{2}+\bar{M}^{2}_{3}+H^{2}\frac{\delta_{4}}{2}+3H^{2}\delta_{3}-\bar{M}_{4}H, \\
F_1 &=& -2M^{2}_{Pl}\dot{H}-\bar{M}^{3}_{1}H-3H^{2}\bar{M}^{2}_{2}-\bar{M}^{2}_{3}H^{2} \nb\\
&& +3H^{4}\left(\delta_{3}+\frac{3}{2}\delta_{4} \right)-\bar{M}_{4}H^{3}.
\eqn
and
\bqn
F_{0}(k,\tau)\equiv\frac{9}{2}\delta_{3}-\frac{27}{4}\delta_{4}-\frac{k^{2}}{2a^{2}H^{2}}(\delta_{3}+3\delta_{4})-\frac{9}{2H}\bar{M}_{4}.  \nb\\
\eqn
In order to simplify the above equation,   assuming $\delta_3=-3\delta_4$ \footnote{The discussion without this condition has been studied in \cite{ashoorioon_nonunitary_2018} recently.} and $u_k=a \pi_k$
\cite{ashoorioon_extended_2018a},  we obtain
\bqn \lb{eom}
u''_k + \left( \omega_k^2(\eta)-\frac{z''}{z}\right)u_k=0,\nb\\
\eqn
where
\bqn\lb{nonlinear}
\omega_k^2(\eta) = c_s^2 k^2 \left(1+\frac{D_1}{G_1 c_s^2} \frac{k^2}{a^2} + \frac{C_1}{G_1 c_s^2} \frac{k^4}{a^4}\right),
\eqn
and
\bqn
c_s^2 &\equiv & \frac{F_{1}}{G_{1}}, \lb{cs}\\
G_{1}&\equiv & -2M^{2}_{Pl}\dot{H}+4M^{4}_{2}-6\bar{M}^{3}_{1}H-9H^{2}\bar{M}^{2}_{2} \nb\\
&&-3H^{2}\bar{M}^{2}_{3}-\frac{81}{2}H^{4}\delta_{4}-9\bar{M}_{4}H^{3}.
\eqn
In the quasi-de-Sitter limit, we also have
\bqn
\frac{z''}{z} \simeq \frac{2}{\eta^2}, \;\; a H \simeq - \eta.
\eqn
It is worth to note that, after considering the high-order derivative terms in the action, the conventional linear dispersion relation becomes nonlinear. To study their  effects it is convenient to introduce a characteristic energy scale $M_*$, above  which the nonlinear terms become dominant. For this purpose, we set
\bqn
\hat \alpha_0 \left(\frac{H}{M_*}\right)^2 = \frac{D_1 H^2}{G_1 c_s^4}, \lb{alpha_0}\\
\hat \beta_0 \left(\frac{H}{M_*}\right)^4 = \frac{C_1}{G_1 c_s^6}, \lb{beta_0}
\eqn
where $\hat \alpha_0$ and $\hat \beta_0$ are two dimensionless constants in the de-Sitter limit. Since the physical  observables are evaluated at the time when the perturbation modes exit the Hubble radius with the energy scale of the order of the Hubble scale $H$, in order for the high-order derivative terms  to be under control, one would require $H \ll M_*$,  so that the horizon crossing can occur in the regime where the dispersion relation acquires the standard linear form. With this in mind, it is convenient to write the equation of motion for the scalar perturbations in the form, 
\bqn \lb{eom_1}
u''_k(\eta) + c_s^2 k^2 \left(1+\hat \alpha_0 \epsilon_*^2  y^2+\hat \beta_0 \epsilon_* ^4  y^4- \frac{2}{y^2}\right)u_k(\eta)=0,\nb\\
\eqn
where we define $y \equiv - c_s k \eta$ and
\bqn\lb{epsilon}
\epsilon_* \equiv \frac{H}{M_*} \ll 1.
\eqn
On the other hand, we also require that all the perturbation modes have to be stable in the ultraviolet regime, which leads to the condition for the healthy ultraviolet behavior
\bqn\lb{uv}
\hat \beta_0 >0.
\eqn

Then, a natural  question arises: whether the scalar perturbations are still compatible with observations after  the inclusion of  high-order operators. It was claimed that the theory with high-order operators would have an IR strong coupling cut-off $\Lambda^{\rm IR}_c$, which  in general makes this theory not a viable EFT. If this is true, it implies that the high energy regime  dominated by the $k^6$ term in (\ref{nonlinear}) is not accessible,  and all the high-order operators are suppressed by the strong coupling cut-off,  which can only contribute some negligible corrections. However, as argued in \cite{ashoorioon_extended_2018a, Baumann:2011su}, this may not always be the case. One way to solve this issue is to require the energy scale of the strong coupling is greater than the energy scale of the new physics. In this way, the high-order operators will dominate at the high energy regime before the scalar perturbation mode becomes strong coupled, which changes dramatically the scaling behavior of the theory,  so that {it can be healthy in both  of the IR and UV limits. Similar treatment has also been employed in studying the strong coupling problem in the HL theory of quantum gravity and plays an essential role for making the theory power-counting renormalizable \cite{horava_general_2010, zhu_symmetry_2011, zhu_general_2012}. With this treatment, the extended EFT of inflation with high-order} operators can provide a controlled description of the perturbations around a quasi de Sitter background from the low energy regime to UV regime \cite{ashoorioon_extended_2018a}). In Appendix A, by considering two cubic and quartic terms in the action as examples, we show in detail how the presence of the high-order operators can cure the strong coupling problem.

In the next two sections, we shall show in detail that it is the behavior of the high-order operators in the UV regime that leads to significant effects on the evolution of the scalar perturbations and produces modifications on the primordial power spectrum. 


\section{Approximate solution in the uniform asymptotic approximation}
\renewcommand{\theequation}{3.\arabic{equation}}\setcounter{equation}{0}

\subsection{WKB Approximation}

In this section, we start with the evolution of the scalar perturbations during the inflation with the nonlinear dispersion relation given in the last section. In general, an important feature of the nonlinear dispersion relation is that it can produce additional excited states for the primordial perturbations on the sub-horizon scale during inflation. Before we study the generation of these excited states and their effects on the 
primordial perturbation spectrum in detail, we would like first to provide a qualitative analysis by using the WKB approximation.

 In general, the solution of the mode function $\mu_k^{(s,t)}(\eta)$ of the equation,
\bqn
\mu_k''^{(s,t)}(\eta)+ \Omega^2(\eta) \mu_k^{(s,t)}(\eta)=0,
\eqn
can be approximated by the WKB solutions
\bqn\lb{eom_wkb}
\mu_k^{(s, t)}(\eta) \simeq \frac{\alpha_k}{\sqrt{2 \Omega(\eta)}}e^{- i \int \Omega(\eta) d\eta} + \frac{\beta_k}{\sqrt{2 \Omega(\eta)}}e^{ i \int \Omega(\eta) d\eta},\nb\\
\eqn
if the WKB condition
\bqn \lb{wkb}
\left|\frac{3 \Omega' {^2}}{4\Omega^4}-\frac{\Omega''}{2\Omega^3}\right|\ll 1,
\eqn
is satisfied. Here the function
\bqn\lb{Omega}
\Omega^2(\eta) &\equiv& \omega_k^2(\eta) - \frac{z''}{z}\nb\\
&=& c_s^2 k^2 \left(1+\hat \alpha_0 \epsilon_*^2  y^2+\hat \beta_0 \epsilon_* ^4  y^4- \frac{2}{y^2}\right),
\eqn
 and $\alpha_k$ and $\beta_k$ are the two Bogoliubov coefficients, which will be determined by {the initial conditions}. Generally an adiabatic state is assumed \cite{baumann_tasi_2009}, 
\bqn
\alpha_k=1, \; \beta_k =0.
\eqn
However, in some cases, the WKB condition may be violated or not be satisfied during the whole process. Then, the non-adiabatic evolution of the mode $\mu_k(\eta)$ will produce excited states (i.e. particle production)  and eventually lead to a state with
\bqn
\alpha_k \neq 1,\; \beta_k \neq 0.
\eqn

\begin{figure}
\includegraphics[totalheight=2.4in,width=3.4in,angle=0]{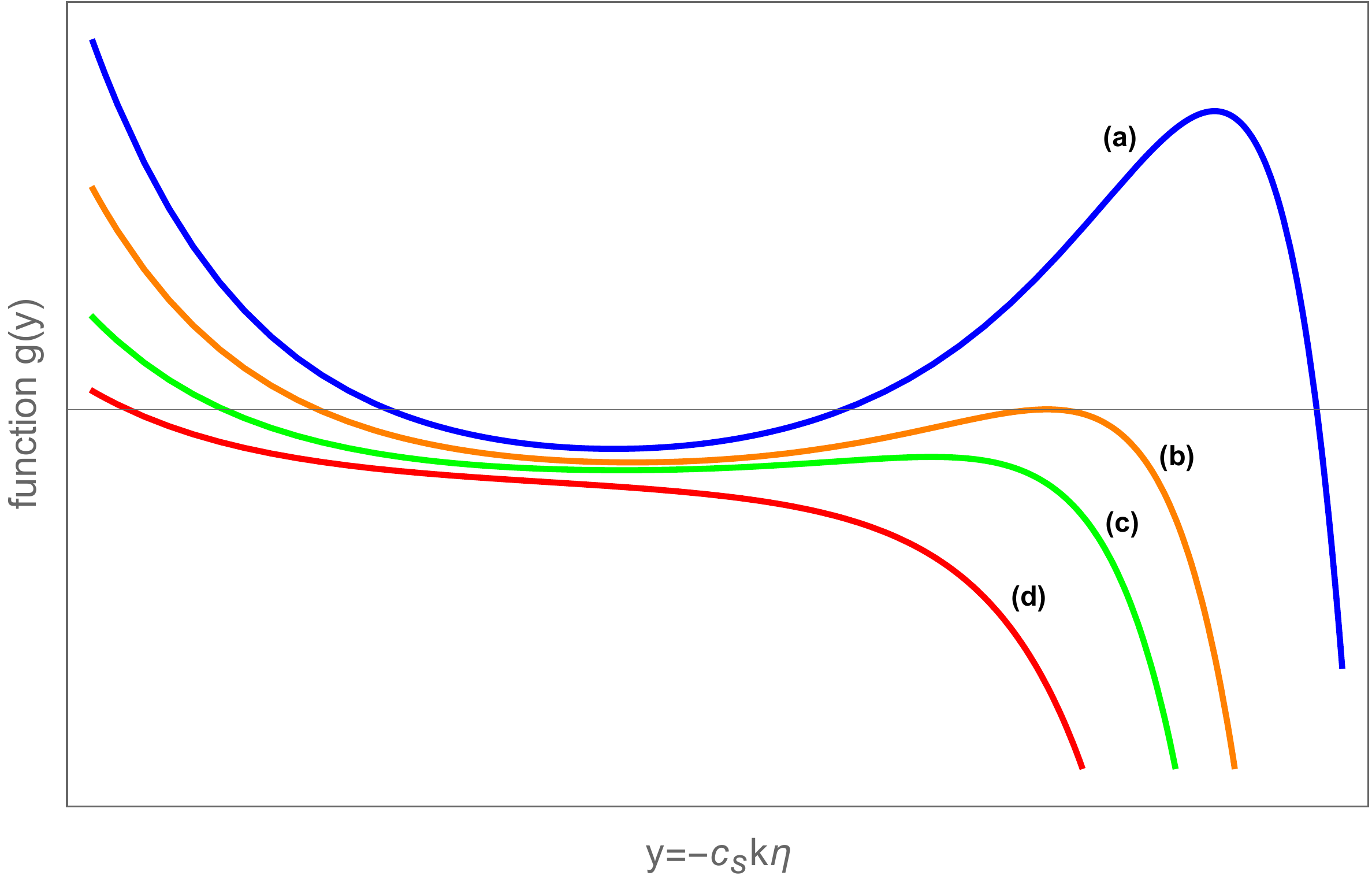}
\caption{The function $g(y)$ defined by Eq.(\ref{G}). The cases (a, b, c) correspond, respectively, to (a) three different real roots; (b) one single and double roots;  or (c) a single and two complex conjugated roots. The case (d) has only a single real root with $\hat{\alpha}_0<0$. In all cases, $\hat{\beta}_0>0$ is assumed. }\label{gofy_tu}
\end{figure}

According to (\ref{wkb}), there are several situations in which the WKB condition can be violated. One case is when $\Omega^2(\eta)$ contains zeros (represented as the real turning points of Eq. (\ref{eom_wkb}) or case (a) and (b) in Fig.~\ref{gofy_tu}) or is extremely close to zero (complex conjugated turning points of Eq.~(\ref{eom_wkb}) or case (c) in Fig.~\ref{gofy_tu}) in the intervals of interest. It is simple to check that when $\Omega^2(\eta)$ equals zero, the WKB condition of (\ref{wkb}) becomes divergent. While for the linear dispersion relation, $\Omega^2(\eta)$ in general has only one zero ( which can be identified as the time when the mode exit the Hubble radius),  $\Omega^2(\eta)$ in Eq.~(\ref{Omega}) may have three zeros because of the presence of the higher order operators in the nonlinear dispersion relation. Then, the WKB condition can be violated at several points. In particular, when the two zeros are real, the WKB condition is strongly violated, while it is only weakly violated if these two zeros are complex conjugated. Another possible case that could violate the WKB condition is around the second-order pole about $y \to 0^+$ (i.e. $\eta \to 0^-$). For the latter, it can be shown that
\bqn
\left|\frac{3 \Omega' {^2}}{4\Omega^4}-\frac{\Omega''}{2\Omega^3}\right|\to \frac{3}{8} \sim \mathcal{O}(1),
\eqn
so the WKB condition is not satisfied. Recently, in order to deal with such cases, we have developed a method (the uniform asymptotic approximation).  In the following subsections, we are going to apply this method to the perturbation modes with the high-order operators and study their effects in detail.

\subsection{Classification of turning points}

To this purpose, let us first write the equation of motion (\ref{eom}) for $u_k$ in the standard form for the {uniform asymptotic approximation \cite{olver_asymptotics_1997, olver_secondorder_1975, zhu_inflationary_2014}}
\bqn
\frac{d^2\mu_k}{dy^2} = \{g(y)+q(y)\} \mu_k,
\eqn
where
\bqn\lb{GQ}
g(y)+q(y)=\frac{2}{y^2}-\hat \beta_0  \epsilon_*^4 y^4- \hat \alpha_0 \epsilon_*^2  y^2-1.
\eqn
According to the theory of the second-order ordinary differential equations, the solution of {the} above equation depends on poles and turning points of the function $g(y)$ and $q(x)$. Analyzing the corresponding error control function associated with the uniform asymptotic approximation around poles and turning points can provide guidance on how to determine the functions $g(y)$ and $q(x)$. For the equation of motion given in the above, we can see that $g(y)$ and $q(y)$ in general has two poles: one is located at $y=0^+$ and the other is at $y=+\infty$. Using the analysis in the uniform asymptotic approximation \cite{olver_asymptotics_1997, olver_secondorder_1975, zhu_inflationary_2014}, the  functions $g(y)$ and $q(y)$ can be chosen as
\bqn\lb{G}
q(y)&=&-\frac{1}{4y^2},\nb\\
g(y)&=&\frac{9}{4y^2}-1-\hat{\alpha}_0 \epsilon^2_\ast y^2-\hat{\beta}_0\epsilon^4_\ast y^4.
\eqn

Except the two poles at $y=0^+$ and $y=+\infty$, $g(y)$ may also have zeros in the range $y\in(0,+\infty)$, which are called turning points. By solving the equation $g(y)=0$, we obtain three turning points, which are
\bqn
y_0&=&\left\{\frac{-\hat{\alpha_0}}{3\hat{\beta_0}\epsilon_*^2}\left[1-2\sqrt{1-Y}\cos\left(\frac{\theta}{3}\right)\right]\right\}^{1/2}, \nb\\
y_1&=&\left\{\frac{-\hat{\alpha_0}}{3\hat{\beta_0}\epsilon_*^2}\left[1-2\sqrt{1-Y}\cos\left(\frac{\theta+2\pi}{3}\right)\right]\right\}^{1/2},\nb\\
y_2&=&\left\{\frac{-\hat{\alpha_0}}{3\hat{\beta_0}\epsilon_*^2}\left[1-2\sqrt{1-Y}\cos\left(\frac{\theta+4\pi}{3}\right)\right]\right\}^{1/2},\nb\\
\eqn
with $Y\equiv3\hat{\beta}_0/\hat{\alpha}_0^2$ and
\bqn
\cos\theta\equiv-\left(1-\frac{3}{2}Y-\frac{27}{8}\hat{\alpha}_0Y^2\epsilon_*^2\right)^2.
\eqn
Without loss of generality we assume that $0<y_0 < {\rm Re}(y_1) \leq {\rm Re}(y_2)$, in which $y_0$ is assumed to be a single and real turning point but $y_1$ and $y_2$ can be both real and single, real and double, or complex conjugated. In general, the nature of $y_1$ and $y_2$ can be determined by,
\bqn
\Delta \equiv(\mathcal{Y}-1)^3+\left(1-\frac{3}{2}\mathcal{Y}-\frac{27}{8}\hat{\alpha}_0Y^2\epsilon_*^2\right)^2.
\eqn
When $\Delta <0$, the three turning points ($y_0$, $y_1$, and $y_2$) are all real and different. When $\Delta =0$, there is one single real turning point ($y_0$), and one double real turning point ($y_1=y_2$). When $\Delta >0$, there is a single real turning point  ($y_0$) and two complex conjugated turning points ($y_1^\ast=y_2$).

\subsection{Approximate solution in the Uniform Asymptotic approximation}

According to the discussions given above, there are two poles and three turning points. In the uniform asymptotic approximation, the approximate solution depends on the types of the turning points. Thus in the following we are going to discuss the solution around each turning point in detail.

We first consider the single turning point $y_0$, which lies in the range $(0, {\rm Re}(y_1)$). Then, the approximate solution around this single turning point can be expressed in terms of the Airy functions,
\bqn\lb{Ar}
\mu_k(y)=a_0\left(\frac{\xi(y)}{g(y)}\right)^{1/4}{\rm Ai}(\xi)+b_0\left(\frac{\xi(y)}{g(y)}\right)^{1/4}{\rm Bi}(\xi),\nb\\
\eqn
where ${\rm Ai}(\xi)$ and ${\rm Bi}(\xi)$ are the Airy functions {of type I and II, respectively}, $a_0$ and $b_0$ are two integration constants, and $\xi$ is a monotonous function of $y$ given by
\bqn
\xi(y) =
\begin{cases}
\left(-\frac{3}{2}\int^y_{y_0}\sqrt{g(y')}dy'\right)^{2/3} ,\;  & 0<y\leq y_0,\\
-\left(\frac{3}{2}\int^y_{y_0}\sqrt{-g(y')}dy'\right)^{2/3} ,\; & y_0<y\leq {\rm Re}(y_1),\\
\end{cases}\nb\\
\eqn.

{Around the the turning points $y_1$ and $y_2$,   the approximate solution of $\mu_k(y)$ can be expressed as}
\bqn\lb{parabolic}
\mu_k(y)&=&a_1\left(\frac{\zeta^2-\zeta^2_0}{-g(y)}\right)^{\frac{1}{4}}W\left(\frac{1}{2}\zeta^2_0,\sqrt{2}\zeta\right)\nb\\
&&+b_1\left(\frac{\zeta^2-\zeta^2_0}{-g(y)}\right)^{\frac{1}{4}}W\left(\frac{1}{2}\zeta^2_0,-\sqrt{2}\zeta\right),
\eqn
where $W(\frac{1}{2}\zeta^2_0,\sqrt{2}\zeta)$ and $W(\frac{1}{2}\zeta^2_0,-\sqrt{2}\zeta)$ are the parabolic cylinder functions, $a_1$ and $b_1$ are two integration constants, and $\zeta_0^2$ is defined as
\bqn
\zeta_0^2 = \pm \frac{2}{\pi} \left|\int_{y_1}^{y_2}\sqrt{g(y)}dy\right|.
\eqn
Here $\pm$ correspond to $y_{1,2}$ being both real and complex conjugated respectively. We observe that the sign of $\zeta_0^2$ depends on the type of the turning points $y_1$ and $y_2$. $\zeta_0^2$ is positive when $y_1$ and $y_2$ are real and negative if $y_1$ and $y_2$ are complex conjugated. The variable $\xi$ is a monotonous increasing function of $y$. When $y_1$ and $y_2$ are both real, $\zeta$ is related to $y$ via
\begin{widetext}
\bqn
\begin{cases}
\int_{y_1}^y\sqrt{-g(y^*)}dy^*=\frac{1}{2}\zeta\sqrt{\zeta^2-\zeta^2_0}+\frac{\zeta^2_0}{2}{\rm arcosh}\left(-\frac{\zeta}{\zeta_0}\right),\;y_0<y<y_1,\\
\int_{y_1}^y\sqrt{-g(y^*)}dy^*=\frac{1}{2}\zeta\sqrt{\zeta^2-\zeta^2_0}+\frac{\zeta^2_0}{2}{\rm arcosh}\left(-\frac{\zeta}{\zeta_0}\right),\;y_1<y<y_2,\\
\int_{y_2}^y\sqrt{-g(y^*)}dy^*=\frac{1}{2}\zeta\sqrt{\zeta^2-\zeta^2_0}-\frac{\zeta^2_0}{2}{\rm arcosh}\left(\frac{\zeta}{\zeta_0}\right),\;y_2<y,
\end{cases}
\eqn
When $y_1$ and $y_2$ are complex conjugated, we have
\bqn
\int_{Re(y_1)}^y\sqrt{-g(y^*)}dy^*=\frac{1}{2}\zeta\sqrt{\zeta^2-\zeta^2_0}-\frac{\zeta^2_0}{2}\ln\left(\frac{\zeta+\sqrt{\zeta^2-\zeta^2_0}}{|\zeta_0|}\right),
\eqn
\end{widetext}

With the approximate solutions around each of the turning points given above, now we need to match them together. Before doing so, 
 we need first to specify the initial conditions of the perturbation modes. As we have mentioned at the end of Sec. II (see Eq.~(\ref{uv})), in order to obtain a healthy ultraviolet limit, we have $\hat \beta_0 >0$. This allows us to impose the usual adiabatic Bunch-Davies vacuum state when $y \to +\infty$,
\bqn
\lim_{y \to +\infty}\mu_k(y)&=&\frac{1}{\sqrt{2\omega_k}}e^{-i\int\omega_k d\eta}\nb\\
&=&\sqrt{\frac{1}{2k}}\left(\frac{1}{-g}\right)^{1/4}\exp\left(-i\int^y_{y_i}\sqrt{-g}dy\right).\nb\\
\eqn
Once the initial conditions are specified, we require the approximate solution (\ref{parabolic})  satisfy these conditions $y \gg {\rm Re} (y_2)$, so that we obtain
\bqn
a_1&=&2^{-3/4}k^{-1/2}\kappa^{-1/2}\left(\frac{1}{2}\zeta_0^2\right),\\
b_1&=&-i2^{-3/4}k^{-1/2}\kappa^{1/2}\left(\frac{1}{2}\zeta_0^2\right),
\eqn
where $\kappa\left(\frac{1}{2}\zeta_0^2\right)$ is given by
\bqn
\kappa\left(\frac{1}{2}\zeta_0^2\right)\equiv\sqrt{1+e^{\pi\zeta^2_0}}-e^{\pi\zeta^2_0/2}.
\eqn
Then we need to match the approximate solution (\ref{Ar}) around the single turning point $y_0$ with the approximate solution (\ref{parabolic}) around the turning point $y_1$ in their overlaping region between $y_0$ and $y_1$. This leads to, 
\bqn\lb{ab}
a_0&=&\sqrt{\frac{\pi}{2k}} \left[\kappa^{-1}\left(\frac{1}{2}\zeta_0^2\right)\sin\mathfrak{B}-i \kappa\left(\frac{1}{2}\zeta_0^2\right)\cos\mathfrak{B} \right],\nb\\
b_0&=&\sqrt{\frac{\pi}{2k}} \left[\kappa^{-1}\left(\frac{1}{2}\zeta_0^2\right)\cos\mathfrak{B}+i \kappa\left(\frac{1}{2}\zeta_0^2\right)\sin\mathfrak{B}\right],\nb\\
\eqn
where
\bqn
\mathfrak{B}\equiv\int^{y_1}_{y_0}\sqrt{-g}dy+\phi(\zeta_0^2/2),
\eqn
with $\phi(x)=\frac{x}{2} - \frac{x}{4}\ln x^2 + \frac{1}{2} {\rm ph} \Gamma(\frac{1}{2}+i x)$, where the phase of $ \Gamma(\frac{1}{2}+i x)$ is zero when $x=0$, and is determined by continuity otherwise.

\begin{figure*}
\centering
\includegraphics[totalheight=2.4in,width=3.4in,angle=0]{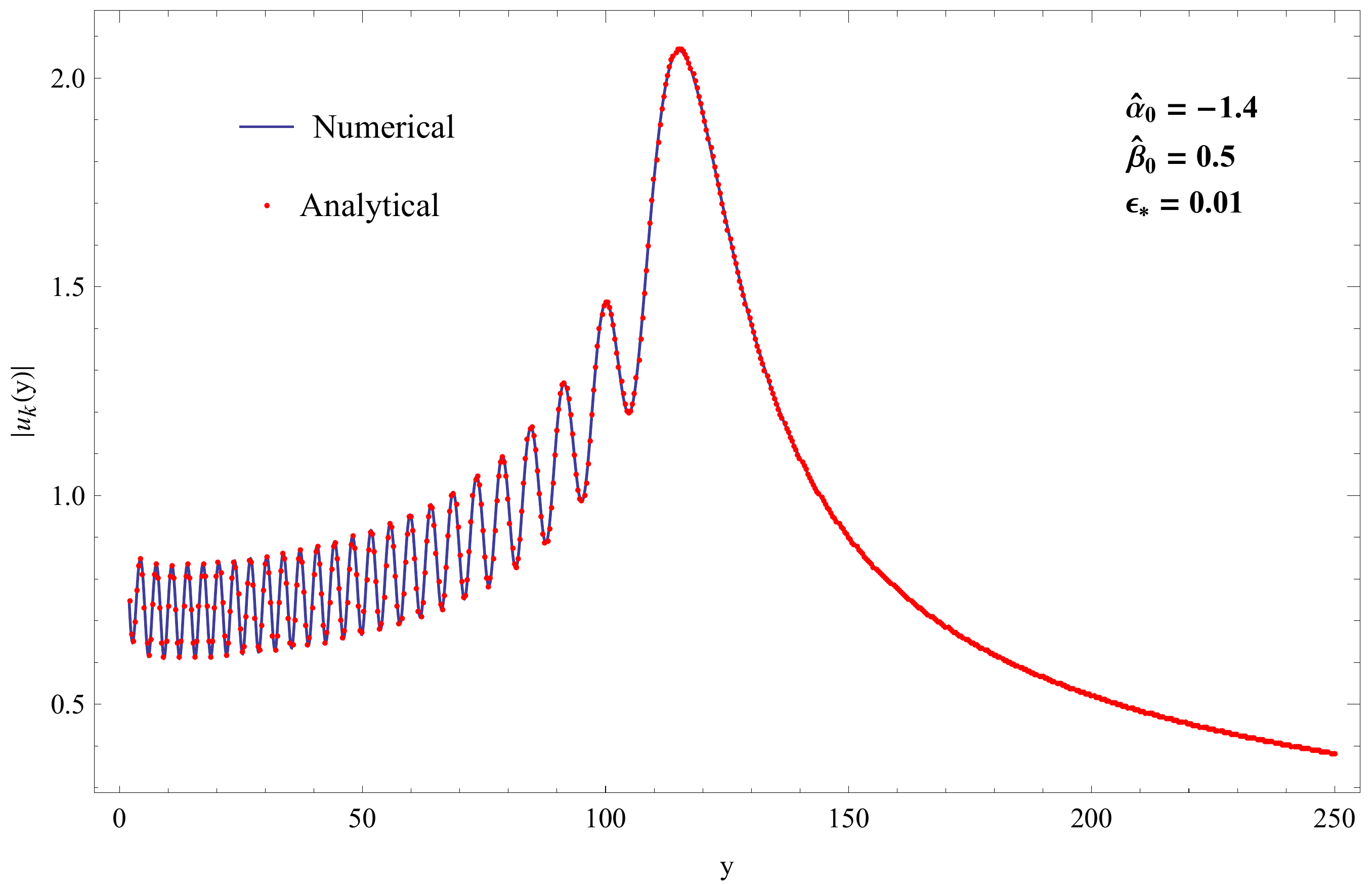}
\includegraphics[totalheight=2.4in,width=3.4in,angle=0]{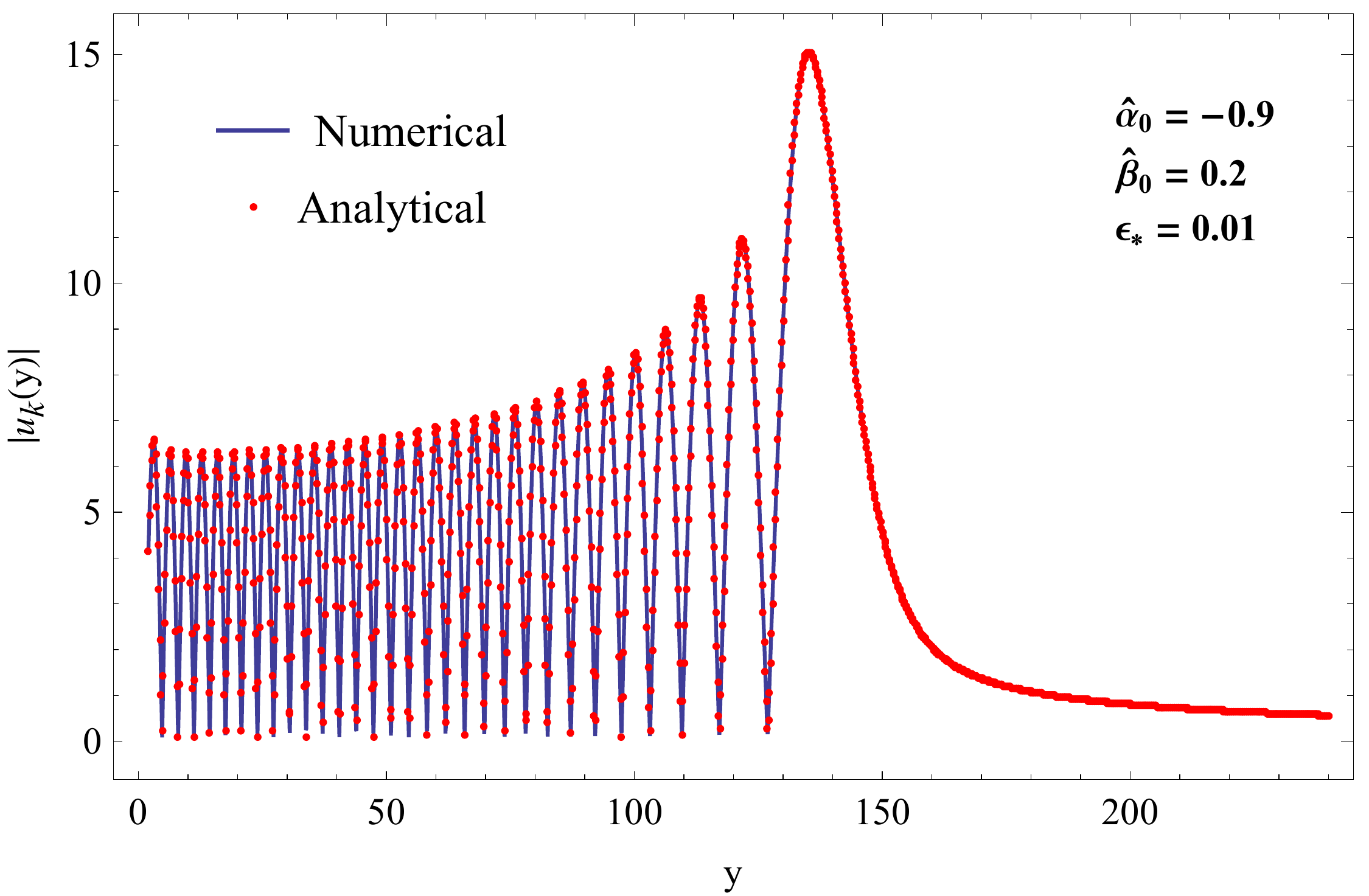}
\caption{Comparison of the analytical approximate solutions in the uniform asymptotic approximation to the numerical (exact) solutions in a de Sitter background.}\label{num_ana}
\end{figure*}

Having matched the approximate solutions together,  all the integration constants appearing in the approximate solutions are uniquely determined by the initial conditions. Therefore, with these solutions, we are able to study the perturbation modes starting from the high energy regime until the end of the slow-roll inflation. Let us now consider some representative cases. The cases with (i) three different single turning points, and (ii) one and two complex conjugate roots, are plotted respectively in the left and right panels of Fig.~\ref{num_ana}. From these figures, we can see clearly that the exact solutions are well approximated by the analytical ones. We have also considered many other cases, and found that   in all the cases the analytical approximate solution traces the exact (numerical) one very well.
 
\section{Non-adiabatic effects on power spectrum of the scalar perturbations}
\renewcommand{\theequation}{4.\arabic{equation}}\setcounter{equation}{0}

As we have mentioned in the above section, the presence of the {extra turning points} leads to the violation of the adiabatic evolution of the perturbation modes. This fact has also been indicated and discussed in detail in Refs.~\cite{ashoorioon_extended_2018, ashoorioon_getting_2017, zhu_inflationary_2014, zhu_highorder_2016}. {As pointed out} in these works, the non-adiabatic evolution of the perturbation modes leads to highly populated excited states and amplify the standard {perturbation spectrum}. Since the presence of the extra turning points is a direct consequence of the high-order operators included in the extended EFT of inflation, these non-adiabatic effects are caused directly by these operators. In this section, by using the analytical solution we have derived above, we discuss in detail how these effects affect both the generation of excited states and the primordial perturbation spectrum.

\subsection{Generation of excited states and particle production rate}

Let us first consider the non-adiabatic effects on the generation of the excited state. For this purpose, when the perturbation modes are inside the Hubble radius (between $y_0$ and $y_1$),  {we find that
 $g(y)\simeq-\omega^2_k/k^2$ and}
\bqn
\int^y_{y_0}\sqrt{-g(y)}dy'\simeq-\int^{\eta}_{\eta_0}\omega_k(\eta)d\eta'.
\eqn
{Then, using the asymptotic form of the Airy functions, we find that the approximate solution (\ref{Ar})  can be casted in the form}
\bqn
\mu_k(\eta) &\simeq& \frac{1}{\sqrt{2\omega_k}}\sqrt{\frac{k}{2\pi}}\frac{1}{i} \Big\{(ia_0-b_0)e^{ i\int^{\eta}_{\eta_0}\omega_k(\eta)d\eta'- i \frac{\pi}{4}} \nb\\
&&+(ia_0+b_0)e^{-i \int^{\eta}_{\eta_0}\omega_k(\eta)d\eta'- i \frac{\pi}{4}} \Big\},
\eqn
{From which} we can immediately identify the Bogoliubov coefficients of the excited modes at the subhorizon scale as
\bqn
|\beta_k|^2&=&\frac{k}{2\pi} |i a_0-b_0|^2\nb\\
&=&\frac{1}{4}(\kappa^2+\kappa^{-2}-2)\nb\\
&=&e^{\pi\xi^2_0}.
\eqn
Here $\beta_k$ is the Bogoliubov coefficient that measures the particle production rate. From the above expression, we can see that $\beta_k^2$ is determined by $\zeta_0^2$, for which we have the following  remarks. First, the sign of $\zeta_0^2$ is sensitive to the nature of the turning points $y_1$ and $y_2$, which can be classified into several classes:
\begin{itemize}
\item When $y_1$ and $y_2$ are both single and real, $\zeta_0^2>0$, which implies that the particle production during the process is exponentially enhanced. As we have shown in Sec. III. B, for this case to happen, one must require the discriminant $\Delta<0$. This corresponds to a requirement on the parameters of the high-order operators.
\item When $y_1$ and $y_2$ are two real and equal, i.e., $y_1=y_2$, we have $\zeta_0^2=0$. Then, we have $\beta_k^2=1$.
\item When $y_1$ and $y_2$ are complex conjugated, i.e., $y_1=y_2^*$, $\zeta_0^2$ is negative. This implies that the particle production is exponentially suppressed.
\end{itemize}

Now an important question arises for the case of the exponentially enhanced particle production rate, namely, whether or not the backreaction of the excited modes is small enough to allow inflation to last long enough. According to the analysis in \cite{lemoine_stressenergy_2001, brandenberger_backreaction_2005}, in order to avoid large backreactions, one has to impose the condition
\bqn
|\beta_k|^2\lesssim 8\pi \frac{H^2_{inf}M^2_{Pl}}{M^4_\ast},
\eqn
where $H_{inf}$ is the energy scale of the inflation, and the Planck 2015 data yield the constraint $H_{\rm inf}/M_{\rm Pl} \leq 3.5 \times 10^{-5}$ \cite{planck_collaboration_planck_2015-4}. Thus, if we take $H_{\rm inf}/M_{\rm Pl} \sim 2 \times 10^{-3} $, we can infer that
\bqn
|\beta_k|^2\lesssim\mathcal{O}(1).
\eqn
Then{, we can obtain}
\bqn
\sqrt{1+|\beta_k|^2}-|\beta_k|\lesssim |\alpha_k+\beta_k| \lesssim|\beta_k|+\sqrt{1+|\beta_k|^2},\nb\\
\eqn
which leads to the constraint on $|\alpha_k+\beta_k|^2 $, that is,
\bqn
3-2\sqrt{2}\lesssim |\alpha_k+\beta_k|^2 \lesssim3+2\sqrt{2}.
\eqn

\subsection{Scalar perturbation spectrum}

With the solutions obtained in the last subsection, we are now able to calculate the perturbation spectrum for the scalar perturbations at the end of the slow-roll inflation in the limit $y \to 0^+$. In this limit, the scalar perturbation is described by the approximate solution (\ref{Ar}) around $y_0$. This solution can also be represented as a linear combination of one growing mode and one decaying mode, in which ${\rm Bi}(\xi)$ is the growing mode and ${\rm Ai}(\xi)$ is the decaying mode. When $y \to 0^+$, only the growing mode is relevant, and then using the asymptotic form of ${\rm Bi}(\xi)$ we find
\bqn
u_k\simeq b_0\left(\frac{1}{\pi^2g(y)}\right)^{1/4}\exp\left(\int^{y_0}_y\sqrt{g(y)}dy\right),
\eqn
where $b_0$ is given by Eq.(\ref{ab}). Then, the scalar power spectrum is given by
\bqn\lb{pw}
\mathcal{P}_s&\equiv&\frac{k^3}{2\pi^2}\left|\frac{u_k(y)}{z}\right|^2\nb\\
&=& \mathcal{A} \left(\frac{k^2y}{9\pi^2z^2}\right)\exp{\left(2\int^{y_0}_y\sqrt{g(y)}dy\right)},
\eqn
where
\bqn\lb{A}
\mathcal{A}&\equiv&\frac{2k|b_0|^2}{\pi}\nb\\
&=&1+2e^{\pi\zeta_0^2}+2e^{\pi\zeta_0^2/2}\sqrt{1+e^{\pi\zeta_0^2}} \cos(2\mathfrak{B}).
\eqn

Obviously, the perturbation spectrum can be modified due to the high-order operators included in the extended EFT of inflation by  two quantities: the modified factor $\mathcal{A}(k)$ and the exponential integration of $\sqrt{-g(y)}$ from $y_0$ to $0$. In order to see this fact clearly, let us first consider the integral with the linear dispersion relation,
\bqn
\mathcal{M}_0 = \exp{\left(2 \int^{y_0}_y \sqrt{\frac{9}{4 y'^2} -1} dy'\right)}.
\eqn
Here {we have $y_0 = \frac{3}{2}$} in the de-Sitter background.  For the nonlinear dispersion relation, this integral {changes to}
\bqn\lb{integral_M}
\mathcal{M} = \exp{\left(2 \int^{y_0}_y \sqrt{g(y')} dy'\right)}.
\eqn
With the above two definitions, we can cast the perturbation {spectrum of (\ref{pw}) into} the form
\bqn
\mathcal{P}_s = \mathcal{A}\times \frac{\mathcal{M}}{\mathcal{M}_0} \times \mathcal{P}^{\rm GR}_s,
\eqn
where $\mathcal{P}^{\rm GR}_s $ denotes the standard nearly {scale-invariant} power-law spectrum in the framework {of} GR.

One essential question related to the enhanced perturbation spectrum is if the effects of $\mathcal{A}$ and $\mathcal{M}/\mathcal{M}_0$ can lead to the violation of the nearly scale-invariance of the scalar spectrum. In fact, as shown in \cite{zhu_highorder_2016}, if we assume that both parameters $\hat \alpha_0$ and $\hat \beta_0$ are varying slowly, then the resulting primordial perturbation spectrum is still nearly scale-invariant. This assumption is correct if one only considers the first-order slow-roll approximation by treating all the slow-roll quantities as constant. In this case, the quantity $\mathcal{A}$ and $\mathcal{M}/\mathcal{M}_0$ do not depend on $k$, and, as a result, it does not contribute significantly  to any  scale-dependence of the spectrum. In this way, as shown in \cite{zhu_inflationary_2014}, the corresponding scalar spectral index can be calculated directly from the power spectrum (\ref{pw}), which is given by
\bqn
n_s  &\equiv& 1+ \frac{d \ln \mathcal{P}_s}{d\ln k}\nb\\
&=& 4 - 2 \lim_{y \to 0}\int_{y}^{y_0} \frac{1-2 \hat \alpha_0 \epsilon_*^2 y'^2+ 3 \hat beta_0 \epsilon_*^4 y'^4}{\sqrt{g(y')}}dy'\nb\\
&=& n_{s}^{\rm GR}.
\eqn
That is, to the first-order approximations of the slow-roll parameters, the power spectrum indices of the scalar perturbations is the same as that given in GR. This indicates that the presence of the high-order operators in the nonlinear dispersion relation can only affect the overall amplitude of the scalar spectrum at this order.

It is still worth noting that the presence of the high-order operators in the nonlinear dispersion relation can contribute to the scale dependence of the power spectrum, if one goes beyond the first-order slow-roll approximation in the {extended} EFT of inflation. Physically this is because the presence of the high-order operators can affect slightly the time when the scalar perturbation modes exit the Hubble radius, which contributes to {the} spectral index at the order of \cite{zhu_quantum_2014, zhu_highorder_2016},
\bqn
\frac{d \ln\mathcal{A} }{d \ln k} \sim \frac{d \ln \mathcal{M}/\mathcal{M}_0}{ d \ln k} \sim \epsilon_*^2 \times \mathcal{O}(\epsilon),
\eqn
where $\epsilon_*$ denotes the slow-roll parameters. 

Therefore, with the presence of the high-order operators, the resulting power spectrum {of the} scalar perturbations is still nearly scale-invariant and the significant effects due to $\mathcal{A}$ and $\mathcal{M}/\mathcal{M}_0$ can only modify the overall amplitude of the power spectrum. In the following we {consider} the effects of $\mathcal{A}$ and $\mathcal{M}/\mathcal{M}_0$ and their properties in detail.

\subsubsection{Non-adiabatic effects on the perturbation spectrum}

The modified factor $\mathcal{A}$ measures the contribution due to {the} presence of the two turning points $y_1$ and $y_2$. For the perturbation modes with a linear dispersion relation in the EFT of inflation, the equation of motion can in general has only one single turning point. Therefore, the modified factor $\mathcal{A}$ represents a direct effect of the presence of high-order operators included in the extended EFT of inflation, namely, the terms with $\bar M_4$, $\delta_3$, and $\delta_4$ in the action (\ref{DeltaS}).  We observe that $\mathcal{A}$ depends on the quantity $\zeta_0^2$, which {is} related to the strength of the violation of {the} adiabatic evolution because of the {two extra turning points}, as we mentioned above. When $\zeta_0^2$ is positive and large, which corresponds to the case in which both $y_1$ and $y_2$ are real (i.e. case (a) in Fig.~\ref{gofy_tu}), we have
\bqn
e^{\pi \zeta_0^2} \gg 1.
\eqn
Then the modified factor {$\mathcal{A}$ reads}
\bqn
\mathcal{A} \simeq 2 e^{\pi \zeta_0^2} \left(1+ \cos{2 \mathfrak{B}}\right),
\eqn
which indicates that {the} power spectrum is exponentially enhanced {in the most part of the} parameter space. When $\zeta_0^2=0$, which corresponds to the double turning point case $y_1=y_2$ (i.e. case (b) in Fig.~\ref{gofy_tu}), we have
\bqn
\mathcal{A} = 3 + 2 \sqrt{2} \cos{2 \mathfrak{B}}.
\eqn
{In} the case in which $\zeta_0^2$ is negatively large, which corresponds to the case in which the turning points $y_1$ and $y_2$ are complex conjugated, we have
\bqn
e^{\pi \zeta_0^2} \ll 1.
\eqn
Thus, the modified factor $\mathcal{A}$ is
\bqn
\mathcal{A} = 1 + 2 e^{\pi \zeta_0^2/2} \cos{2 \mathfrak{B}}+ \mathcal{O}(e^{\pi \zeta_0^2}),
\eqn
which reduces to the {standard  one} with only one single turning point $y_0$. In Fig.~\ref{modified_factor} and \ref{modified_factor2}, we plot the behavior of the modified factor $\mathcal{A}$ with respect to different parameters. All these figures show clearly the modified factor $\mathcal{A}$ gets enhanced when $y_1$ and $y_2$ are both real and single, and reduces to one {when}  $y_1$ and $y_2$ are complex conjugated. Our analytical results presented here are also in agreement with the numerical results obtained in \cite{ashoorioon_extended_2018}.

\subsubsection{Impact of the exponential integration of $\sqrt{-g(y)}$}

Another effect of the high-order operators, which may also change the exponential integral of $\sqrt{-g(y)}$ over range from $y_0$ to $0$ in comparison to a linear dispersion relation, is the ratio $\mathcal{M}/\mathcal{M}_0$. From Eq.~(\ref{integral_M}), we can see that $\mathcal{M}/\mathcal{M}_0$ is more sensitive to the magnitude of the parameter $\epsilon_* = \frac{H}{M_*}$ rather than the parameters $\hat \alpha_0$ and $\hat \beta_0$. When $\epsilon_* \ll 1$, as we have assumed in Eq.~(\ref{epsilon}), we find the turning point $y_0$ can be approximated by
\bqn
y_0 = \frac{3}{2}+\mathcal{O}(\epsilon_*).
\eqn
During the interval between $0$ and $y_0$, we also have
\bqn
\sqrt{g(y)} = \sqrt{\frac{9}{4 y^2} -1} + \mathcal{O}(\epsilon_*).
\eqn
{With the above expressions,} we have
\bqn
\frac{\mathcal{M}}{\mathcal{M}_0} = 1+ \mathcal{O}(\epsilon_*).
\eqn
This indicates that if $\epsilon_* \ll 1$, the exponential integral in (\ref{integral_M}) does not lead to any significant effects on the primordial perturbation spectrum. The only effects come from the modified factor $\mathcal{A}$.

If we relax the condition $\epsilon_* \ll 1$ to $\epsilon \lesssim \mathcal{O}(1)$, namely, we consider $\hat \alpha_0 \epsilon_*^2 \lesssim \mathcal{O}(1)$ and $\hat \beta_0 \epsilon_*^4 \lesssim \mathcal{O}(1)$, the above conclusion will be {significantly} changed. To see this clearly, in Fig.~\ref{M} we plot $g(y)$ for different values of {the} parameters and compare it with the one of {a linear dispersion relation}. When both $\hat{\alpha}_0\epsilon_\ast^2$ and $\hat{\beta}_0\epsilon_\ast^4$ are positive, which corresponds to case (d) in Fig.~\ref{gofy_tu}, $g(y)$ only has one single turning point and the corresponding modified factor $\mathcal{A} \simeq 1$.  However, as shown in the top panel of Fig.~\ref{M}, this case leads to a shift of $y_0$ from $\frac{3}{2}$ in the linear case to a smaller value. As a result, the curve of $g(y)$ with {a} nonlinear dispersion relation is always beneath the one with {a} linear dispersion relation between $y_0$ and $0$. In this case, {it is obvious  that} $\mathcal{M}/\mathcal{M}_0 <1$ is significant, which implies the perturbation spectrum is suppressed in comparison to the standard one {given in GR.}

When $\hat \alpha_0 \epsilon_*^2$ is negative, things become more complicated. In this case, the shift of the turning point $y_0$ from $\frac{3}{2}$ is a result of competition between $\hat \alpha_0 \epsilon_*^2$ and $\hat \beta_0 \epsilon_*^4$. The former can make $y_0>\frac{3}{2}$ and the later can make it smaller. When the effect of $\hat \alpha_0 \epsilon_*^2$ is larger than that of $\hat \beta_0 \epsilon_*^4$, {then} $y_0$ becomes larger than $3/2$, otherwise it will be smaller. However, from the shift of $y_0$ itself we still cannot conclude if the ratio $\mathcal{M}/\mathcal{M}_0$ is larger than unity or not. As shown in the middle and bottom panels of Fig.~\ref{M}, even when $y_0<3/2$, the curve of $g(y)$ is not always beneath the one with a linear dispersion relation during $(0, y_0)$. This makes the analysis of the ratio $\mathcal{M}/\mathcal{M}_0$  very difficult. However, it can be shown that  the ratio $\mathcal{M}/\mathcal{M}_0$   can be either larger than one or smaller than one, depending on the values of the parameters.

In fact, the impact of the high-order operators on the perturbation spectrum with larger values of $\hat \alpha_0 \epsilon_*^2$ and $\hat \beta_0 \epsilon_*^4$ has already been studied numerically in \cite{ashoorioon_extended_2018}. The results we presented in this subsubsection is in agreement with these numerical analyses. Here we would like to emphasize that our analytical results show clearly the impact of the high-order operators of the extended EFT of inflation on the perturbation spectrum originally from two different effects, one is measured by the modified factor $\mathcal{A}$, and the other is measured by the ratio $\mathcal{M}/\mathcal{M}_0$. In addition, we note that, the suppression or enhancement on the perturbation spectrum due to the ratio $\mathcal{M}/\mathcal{M}_0$ is not an effect of the excited state generated by the non-adiabatic evolution of modes around $y_1$ and $y_2$. These effects mainly originate from the shifts of the turning point $y_0$ in comparison  with the linear dispersion relation.

\begin{figure*}
\centering
\includegraphics[width=8cm]{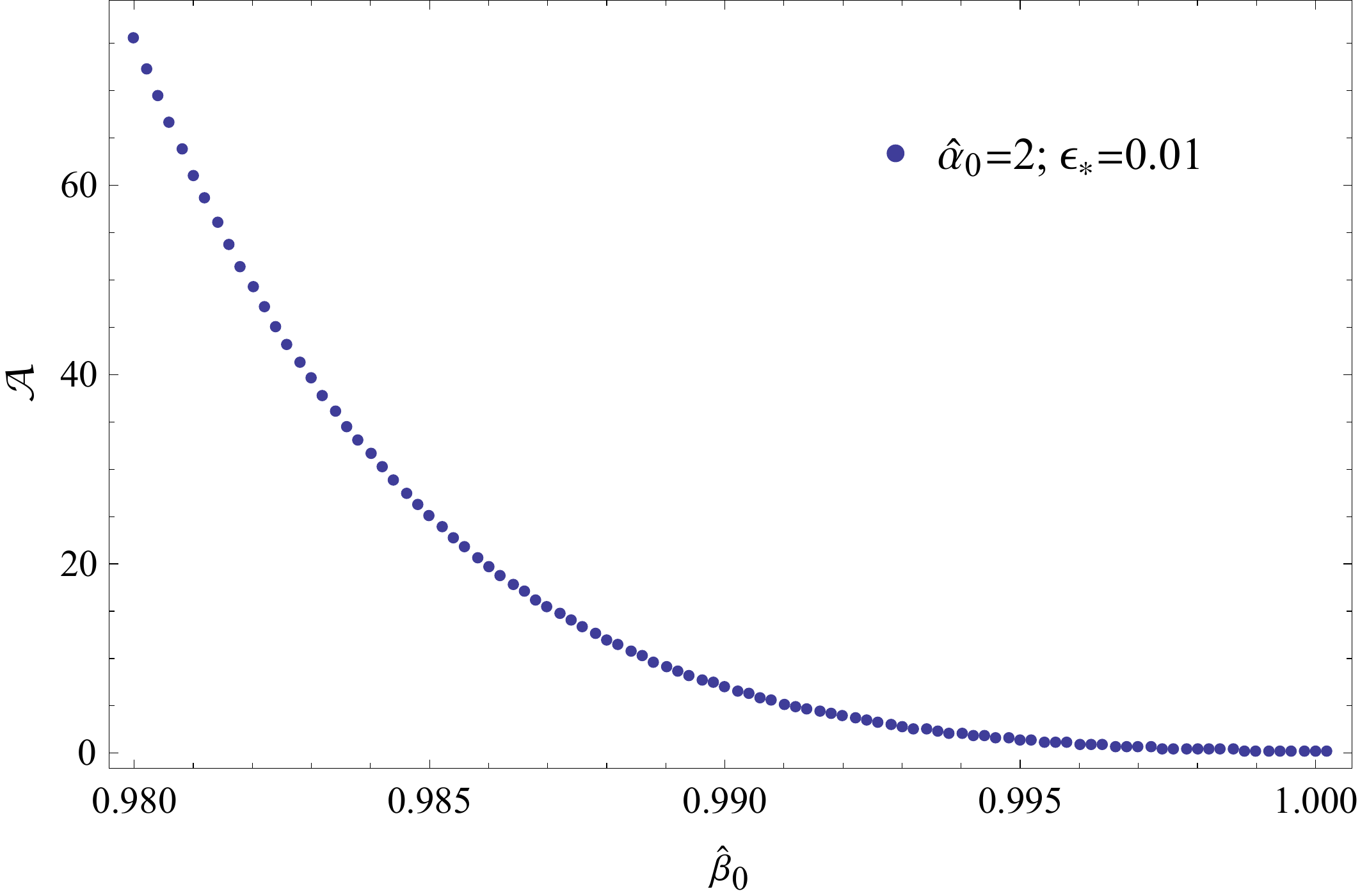}
\includegraphics[width=8cm]{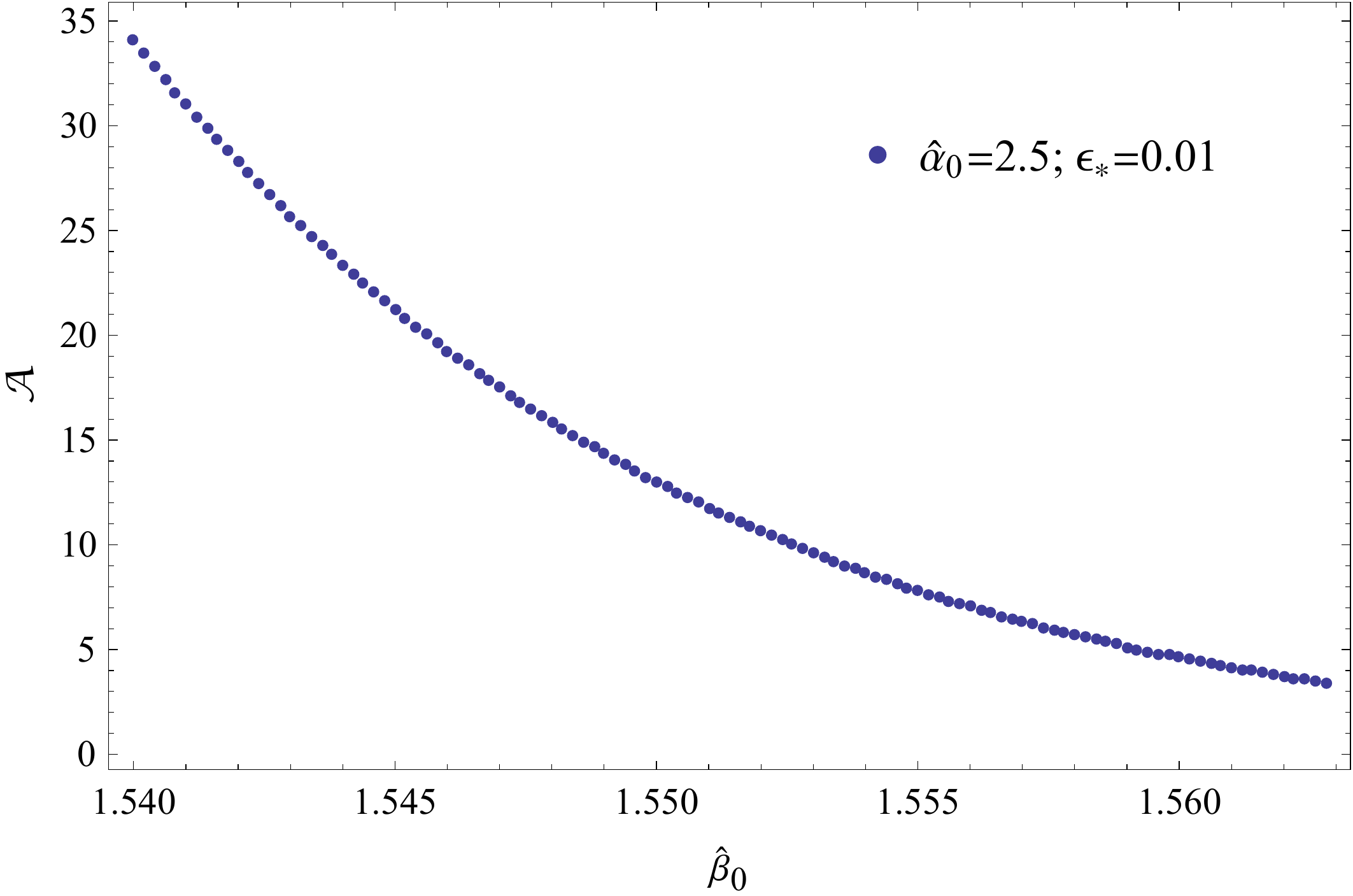}
\caption{ The modified factor $\mathcal{A}$ as a function of $\hat \beta_0$ with $\hat \alpha_0$ and $\epsilon_*$ fixed. In both figures, the modified factor changes from the two real turning points region (corresponds to $\mathcal{A} \gg 1$) to the two complex conjugated region (corresponds to $\mathcal{A} \simeq 1$). Left panel: $\hat \alpha_0 =2$ and $\epsilon_* = 0.01$. Right panel: $\hat \alpha_0 =2.5$ and $\epsilon_* = 0.01$.}
\lb{modified_factor}
\end{figure*}

\begin{figure}
\centering
\includegraphics[width=8cm]{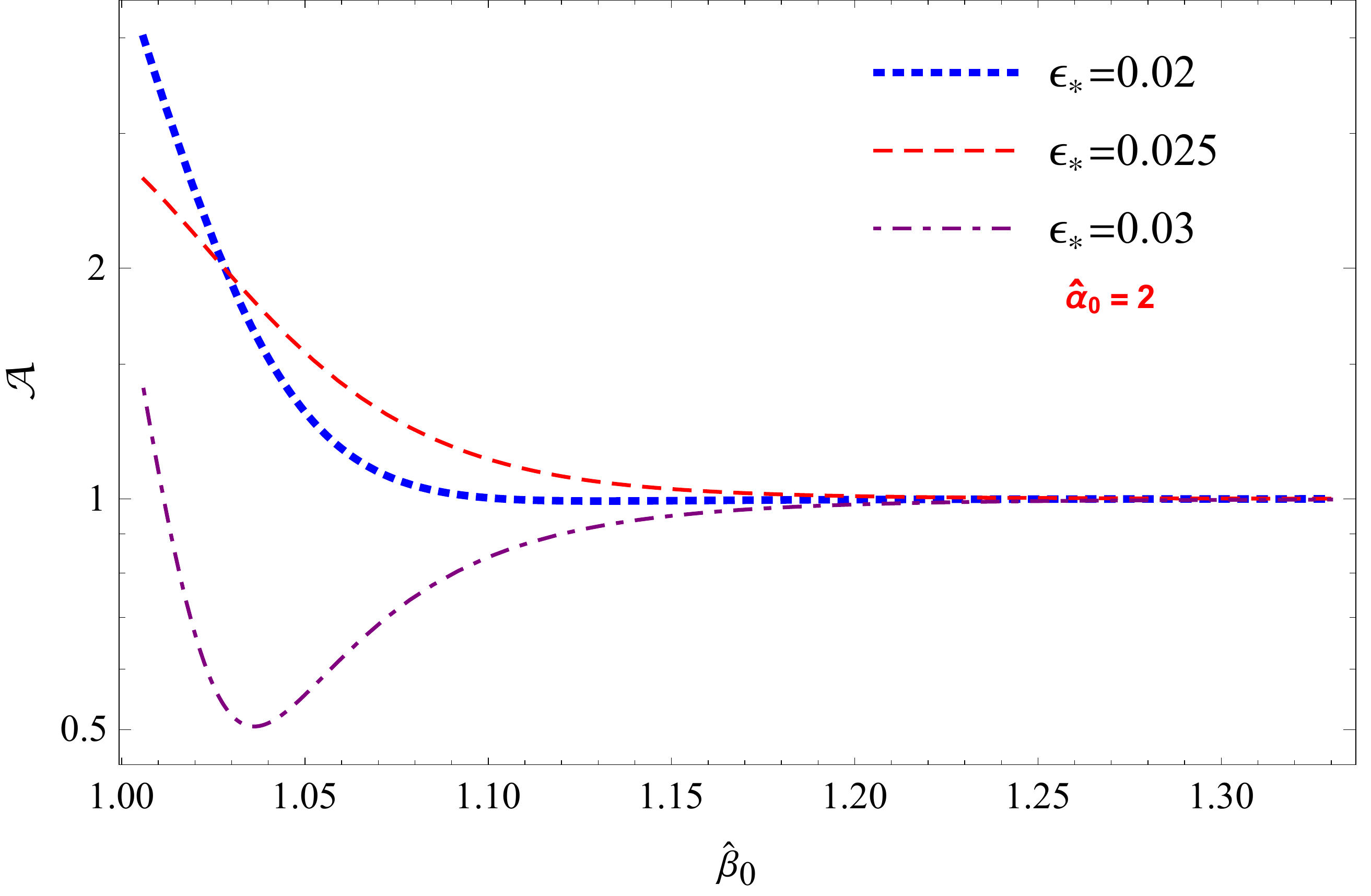}
\caption{The modified factor $\mathcal{A}$ as a function of $\hat \beta_0$ with $\hat \alpha_0=2$ for several values of $\epsilon_*$. }
\lb{modified_factor2}
\end{figure}

\begin{figure}
\centering
\includegraphics[width=8cm]{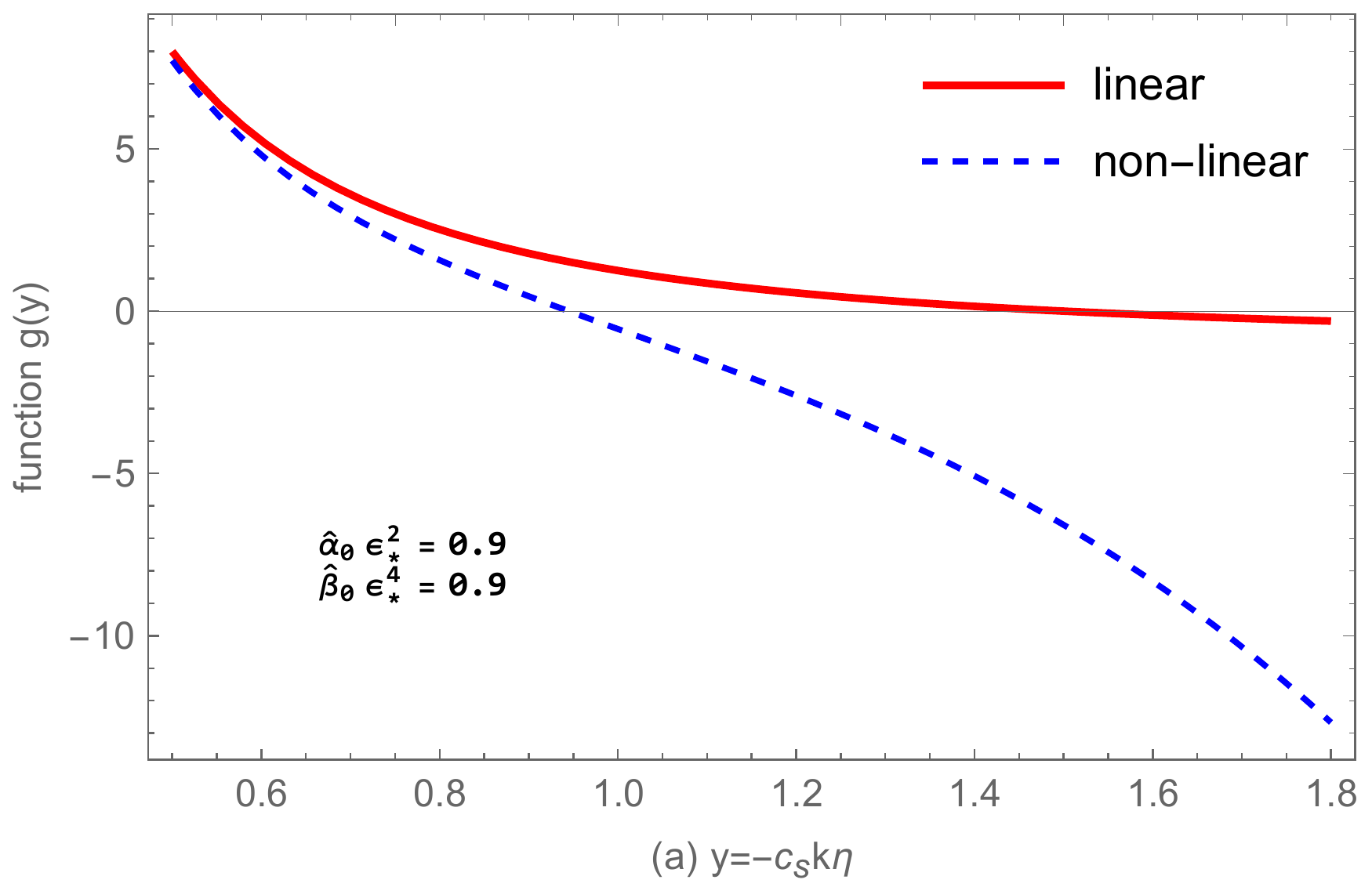}
\includegraphics[width=8cm]{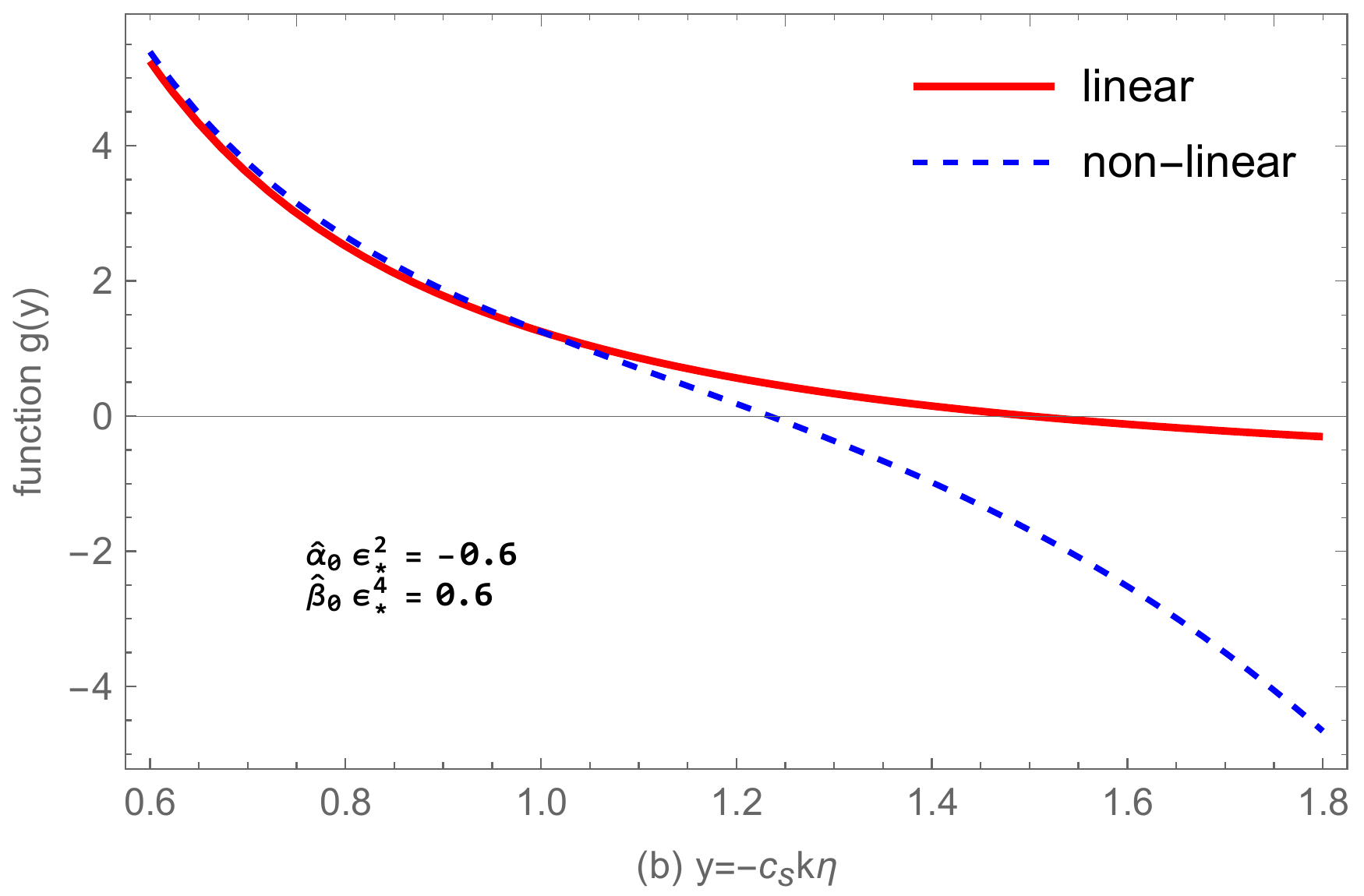}
\includegraphics[width=8cm]{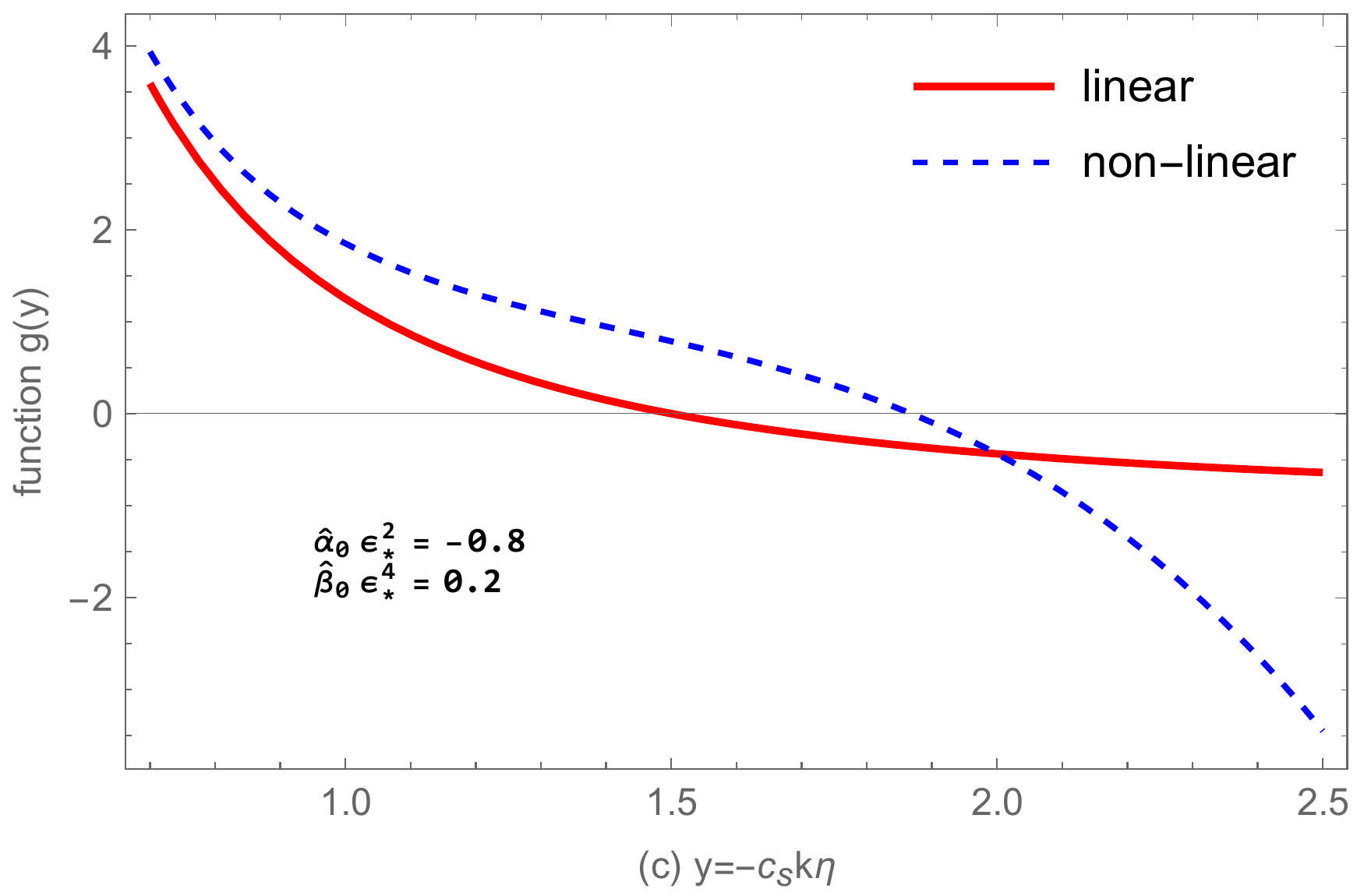}
\caption{Comparison of $g(y)$ with nonlinear and linear dispersion relations in the range of from $(0, y_0)$ for different values of parameters.}\lb{M}
\end{figure}

\section{Conclusion and Outlook}
\renewcommand{\theequation}{5.\arabic{equation}}\setcounter{equation}{0}

In this paper, we {provide} a detailed and analytical study of the effects of high-order operators on the primordial curvature perturbations. These high-order operators are naturally included  in the framework of the extended EFT of inflation. We show clearly that the effects of these high-order operators can naturally generate an excited state on the primordial  curvature perturbation rather than the usual BD vacuum state at subhorizon scale before the perturbation modes cross the Hubble radius. As we showed in Sec. IV, this excited state can produce enhanced effects if these high-order operators can lead to strong violation of the adiabatical condition of the perturbation modes (corresponding to two extra real and single turning points $y_1$ and $y_2$), but reduce to the BD state when the violation is weak (corresponding to two complex conjugated turning points). With this excited state, we calculate explicitly these non-adiabatic effects on  the primordial curvature perturbation spectrum and {showed} that the {modifications} of the spectrum come from two effects, one is from the modified factor $\mathcal{A}$ and the other is from the exponential integral of $\sqrt{g(y)}$ over the range of $(0, y_0)$.  {Despite of all these effects, the} resulting modified power spectrum is still nearly scale-invariant and the presence of the high-order operators can only affect the overall amplitude of the spectrum. In particular, we show clearly that the modified factor $\mathcal{A}$ measures the contribution due to the presence of the two turning points $y_1$ and $y_2$, which represents direct effects of the inclusion of the high-order operators in the theory. The effects from the exponential integral, which either leads to suppression or enhancement on the perturbation spectrum, are mainly  from the derivation of the turning point $y_0$ from that in comparison with the linear dispersion relation.

\section*{Acknowledgements}
This work is supported by National Natural Science Foundation of China with the Grants Nos. 11675143, 11975203, 11675145, and the Fundamental Research Funds for the Provincial Universities of Zhejiang in China with Grants No. RF-A2019015.

\section*{Appendix A: Analysis of the strong coupling problem}
\renewcommand{\theequation}{A.\arabic{equation}}\setcounter{equation}{0}

In this appendix, we present a detailed analysis of the strong coupling problem in the extended EFT of inflation. We consider several cubic and quartic terms in the action which can run into the strong coupling regime when the dispersion relation is linear.  Then we show in detail that this strong coupling problem can be cured by the presence of the high-order operators introduced in the extended EFT of inflation \cite{ashoorioon_extended_2018}. For this purpose, we adopt the mechanism used in refs. \cite{zhu_symmetry_2011, zhu_general_2012, blas_comment_2010, lin_strong_2011} in the Horava-Lifshitz gravity and follow the calculations given there. It is worth noting that the similar mechanism has also been used to get rid of the strong coupling problem in the EFT of inflation with the presence of the fourth-order derivative operators \cite{Baumann:2011su}. 

To proceed, let us first write the quadratic action (\ref{quadra_eff}) of the extended EFT of inflation in the form of
\bqn\lb{quadratic_action}
S_{\pi}^{(2)} &=& \frac{1}{2}\int d^4x a^3 A_1 \Bigg[ \dot \pi^2 -c_s^2 \Bigg( \frac{(\partial_i \pi )^2}{a^2}-\frac{\hat \alpha_0}{M_*^2} c_s^2 \frac{(\partial^2\pi)^2}{a^4} \nb\\
&&~~~~~~~~~~~~~~~~  + \frac{\hat \beta_0}{M_*^4} c_s^4 \frac{(\partial^3 \pi)^2}{a^6}\Bigg) \Bigg],
\eqn
where $A_1$, $c_s$, $\hat \alpha_0$, $\hat \beta_0$, and $M_*$ are given by eqs.~(\ref{A1}, \ref{cs}, \ref{alpha_0}, \ref{beta_0}), respectively.  In order to consider the strong coupling problem, we consider the following cubic and quartic terms in the action {as examples,}
\bqn
S_\pi^{(3)} \sim  \int d^4 x a^3  \lambda_1 \frac{\dot \pi (\partial_i \pi)^2}{a^2},\\
S_{\pi}^{(4)} \sim \int d^4x  a^3 \lambda_2 \frac{(\partial_i \pi \partial^i \pi)^2}{a^4},
\eqn
where $\lambda_1$ and $\lambda_2$ are {two coupling constants}. The coupling constant $\lambda_1$ has been derived in \cite{Baumann:2011su} without the introducing of the sixth derivative operators. The effects of the higher derivative terms can only contribute small corrections so that we have $\lambda_1 \simeq 2 M_{\rm Pl}^2 \dot H (1-c_s^2) c_s^{-2}(1 + \mathcal{O}(\epsilon_*^2))\simeq - 2 M_{\rm Pl}^2 H^2 \epsilon_1 c_s^{-2}(1-c_s^2)$ \cite{Baumann:2011su}. For the coupling constant $\lambda_2$, normally one has $\lambda_2 \sim M_{\rm Pl}^2 H^2 \epsilon_1 (1-c_s^2)^{p}c_s^{-q} (1+\mathcal{O}(\epsilon_*^2))$ with $p \geq 0 , q >0$ \cite{Baumann:2011su}. 

Depending on the energy scale, the above two terms have different scalings. So in the following we consider them, separately.

\subsection{$|\partial_i|  \ll M_*$}

In this case, we find that the high-order derivative terms in the quadratic action can be neglected, and
\bqn\lb{low_action}
S^{(2)}_\pi \simeq \frac{1}{2} \int d^4x a^3 A_1 \left[\dot \pi^2 - c_s^2 \frac{ (\partial_i \pi)^2}{a^2}\right].
\eqn
Considering the transformation 
\bqn \lb{transA}
t \to b_1 \hat t,\;\; x^i \to b_2 \hat x^i,\;\; \pi \to b_3 \hat \pi,
\eqn
the action (\ref{low_action}) can be written in the canonical form,
\bqn\lb{low_actionb}
S^{(2)}_\pi \simeq \int d^4 \hat x a^3  \left[(\hat \partial_t \hat \pi)^2 - \frac{ (\hat \partial_i \hat \pi)^2}{a^2}\right].
\eqn
Note that we consider $A_1$ and $c_s^2$ being constants since they both are slowly-varying during the slow-roll inflation. In writing the above form one sees that the coefficient of each term in the action is of the order of $\mathcal{O}(1)$ and
\bqn
b_2 = b_1 c_s , \;\;\;\; b_3= \frac{\sqrt{2}}{b_1 \sqrt{A_1} c_s^{3/2}}.
\eqn
Then, under the same transformation (\ref{transA}) the cubic and quartic terms transform into
\bqn
S^{(3)}_\pi \simeq \frac{1}{b_1^2 c_s^{7/2}}\left(\frac{2}{A_1}\right)^{3/2} \hat S^{(3)}_\pi,\\
S^{(4)}_\pi \simeq \frac{4}{b_1^4 A_1^2 c_s^7} \hat S_\pi^{(4)}.
\eqn

On the other hand, we observe that the action (\ref{low_actionb}) is invariant under the rescaling,
\bqn
\hat t \to b^{-1} \hat t, \;\; \hat x^i \to b^{-1} \hat x^i,\;\; \hat \pi \to b \hat \pi.
\eqn
Then the coupling constants $\lambda_1$ and $\lambda_2$ scale as $b^2$ and $b^4$, respectively. Therefore, these terms are irrelevant and nonrenormalizable \cite{Polchinski}. To see the problem clearly, let us consider a physical process with the energy $E$, for example, then we have
\bqn
\int d^4 \hat x a^3 \frac{\hat \partial_{\hat t}\hat\pi (\hat \partial_i \hat\pi)^2}{a^2} \sim E^2,\\
\int d^4 \hat x a^3 \frac{ (\hat \partial_i \hat\pi \hat \partial^i \hat \pi)^2}{a^4} \sim E^4.
\eqn
Since the cubic and quartic action is dimensionless, we must have
\bqn
\frac{1}{b_1^2 c_s^{7/2}}\left(\frac{2}{A_1}\right)^{3/2} \hat S^{(3)}_\pi \simeq \left(\frac{E}{\Lambda_{\rm sc}^{(3)}}\right)^2,\\
\frac{4}{b_1^4 A_1^2 c_s^7} \hat S_\pi^{(4)} \simeq \left(\frac{E}{\Lambda_{\rm sc}^{(4)}}\right)^4,
\eqn
where the strong coupling energy scales $\Lambda_{\rm sc}^{(3,4)}$ are,
\bqn
\Lambda_{\rm sc}^{(3)} = \frac{b_1 c_s^{7/4}}{\lambda_1^{1/2}}\left(\frac{A_1}{2}\right)^{3/4}, \\
\Lambda_{\rm sc}^{(4)} = \frac{b_1 \sqrt{A_1}c_s^{7/4}}{\sqrt{2} \lambda_2^{1/4}}.
\eqn
Although we have only considered two terms, by {following} the above procedure one can in principle get the strong coupling scales $\Lambda_{\rm sc}$ for all the nonrenormalizable terms in the cubic and quartic action \cite{Polchinski}. It is obvious that when the energy $E$ is above the strong coupling scales $\Lambda_{\rm sc}^{(3)}$ ($\Lambda_{\rm sc}^{(4)}$), {it runs into} the strong coupling regime. The strong coupling energy and momentum scales in the physical coordinates are given respectively by
\bqn
(\Lambda_{\omega}^{(3)},\Lambda_{k}^{(3)}) = \left(\frac{c_s^{7/4}}{\lambda_1^{1/2}}\left(\frac{A_1}{2}\right)^{3/4}, \frac{c_s^{3/4}}{\lambda_1^{1/2}}\left(\frac{A_1}{2}\right)^{3/4}\right),\nb\\
(\Lambda_{\omega}^{(4)},\Lambda_{k}^{(4)}) = \left(\frac{ \sqrt{A_1}c_s^{7/4}}{\sqrt{2} \lambda_2^{1/4}}, \frac{ \sqrt{A_1}c_s^{3/4}}{\sqrt{2} \lambda_2^{1/4}}\right).\nb\\
\eqn
We would like to mention that the above analysis only holds for $M_* \gg E > \Lambda_{\omega}^{(3, 4)}$, which implies 
\bqn
 \frac{c_s^{7/4}}{\lambda_1^{1/2}}\left(\frac{A_1}{2}\right)^{3/4}, \; \frac{ \sqrt{A_1}c_s^{7/4}}{\sqrt{2} \lambda_2^{1/4}} < M_*.
\eqn
However, when $M_* < \Lambda_\omega^{(3, 4)}$ and $E > M_*$, before reaching the strong coupling energy scale, one has to take the effects of  the high-order derivative terms in (\ref{quadratic_action}) into account. In ref.\cite{Baumann:2011su}, it has been proved in detail that exactly due to the presence of the fourth-order derivative terms in the EFT of inflation, the strong coupling problem can be cured. This is very similar to the strong coupling problem that have been eliminated due to the presence of the high-order derivative terms up to the sixth order \cite{zhu_symmetry_2011, zhu_general_2012}. In the following, by applying the same mechanism we shall show that the strong coupling problem can also be cured with the presence of the high-order spatial derivative terms. Since ref.~\cite{Baumann:2011su} has considered the fourth-order derivative terms in detail, here we only focus on the effects of the sixth-order operators.

\subsection{$M_* < \Lambda_{\omega}^{(3,4)}$}

In this case, for $E > M_*$, the action (\ref{quadratic_action}) reduces to
\bqn
S_{\pi}^{(2)} \simeq \frac{1}{2}\int d^4x a^3 A_1 \Bigg[ \dot \pi^2 - \frac{\hat \beta_0}{M_*^4} c_s^6 \frac{(\partial^3 \pi)^2}{a^6}\Bigg].
\eqn 
Using the transformation (\ref{transA}) with
\bqn
b_2= \frac{2^{1/6}b_1^{1/3} c_s^{1/2} \hat \beta_0^{1/12}}{A_1^{1/6}M_*^{1/3}}, \\
b_3=\frac{2^{1/4}M_*^{1/2}}{A_1^{1/4} c_s^{3/4} \hat \beta_0^{1/8}},
\eqn
one has
\bqn
S_\pi^{(2)} = \int d^4\hat x [(\hat \partial_{\hat t} \hat \pi)^2 - (\hat \partial^3 \hat \pi)^2].
\eqn
For the cubic and quartic action, they transform as
\bqn
S_{\pi}^{(3)} = \frac{2 \sqrt{2} b_1^{1/3} M_*^{7/3}}{A_1^{3/2} c_s^{7/2} \hat \beta_0^{7/12} }\hat S^{(3)}_\pi,\\
S_{\pi}^{(4)} = \frac{ 4 b_1^{2/3} M_*^{14/3}}{A_1^2 c_s^7  \hat \beta_0^{7/6}}\hat S^{(4)}_\pi.
\eqn
Similar to \cite{zhu_symmetry_2011, zhu_general_2012}, we also observe that the quadratic action is invariant under the rescaling
\bqn
\hat t \to b^{-3} \hat t,\;\; \hat x^i \to b^{-1} \hat x^i,\;\; \hat \pi \to \hat \pi.
\eqn
Then, one can immediately see that the $\lambda_1$ term in the cubic action and $\lambda_2$ term in the quartic action scale as $b^{-1}$ and $b^{-2}$, respectively. Therefore, these two terms become
 superrenormalizable \cite{Polchinski}. 

Therefore, the conditions for curing the strong coupling problem which arises in the cubic and quartic terms require $M_* \lesssim \Lambda_{\omega}^{(3,4)}$. This leads to
\bqn\lb{CA}
M_* \lesssim \frac{c_s^{7/4}}{\lambda_1^{1/2}}\left(\frac{A_1}{2}\right)^{3/4}, \; \frac{ \sqrt{A_1}c_s^{7/4}}{\sqrt{2} \lambda_2^{1/4}}.
\eqn
Considering $A_1 \simeq - 2 M_{\rm Pl}^2 \dot H (1+\mathcal{O}(\epsilon_*^2)) \simeq 2 M_{\rm Pl}^2 H^2 \epsilon_1$ and $\lambda_1 \simeq 2 M_{\rm Pl}^2 \dot H (1-c_s^2) c_s^{-2}(1 + \mathcal{O}(\epsilon_*^2))\simeq - 2 M_{\rm Pl}^2 H^2 \epsilon_1 c_s^{-2}(1-c_s^2)$ \cite{Baumann:2011su},  the first condition in (\ref{CA}) reduces to
\bqn
M_*  \lesssim \frac{c_s^{11/4} M_{\rm Pl} ^{1/2}H^{1/2} \epsilon_1^{1/4}}{\sqrt{1-c_s^2}} \simeq \frac{\mathcal{O}(1)}{\sqrt{1-c_s^2}} \times 10^{-3} M_{\rm Pl}.\nb\\
\eqn
Note that in the above we have assumed $c_s \sim \mathcal{O}(1)$ and used $H \sim 2.7 \times 10^{-5} M_{\rm Pl}$, $\epsilon_1 \sim 0.0068$ from the Planck 2018 data \cite{planckcollaboration_planck_2018}. Notice that we have also assumed the condition $H \lesssim M_*$ (i.e. $\epsilon_* < 1$) throughout the paper, which implies 
\bqn
2.7 \times 10^{-5} \lesssim \frac{M_*}{M_{\rm Pl}} \lesssim \frac{ \mathcal{O}(1)}{\sqrt{1-c_s^2}} \times 10^{-3}.
\eqn
Similarly,  for the second condition in (\ref{CA}) with $\lambda_2 \sim M_{\rm Pl}^2 H^2 \epsilon_1 c_s^{-q} (1-c_s^2)^p$,
we find
\bqn
2.7 \times 10^{-5}\lesssim \frac{M_* }{M_{\rm Pl}} \lesssim \frac{\mathcal{O}(1)}{(1-c_s^2)^{p/4}} \times 10^{-3}.
\eqn
Therefore, we conclude that, provided above conditions hold, the cubic and quartic terms considered in this appendix do not lead to any strong coupling problem.  Although we have only considered the strong coupling problem associated with these two terms,  the analysis can be extended to other high-order terms in the cubic and quartic action. { In this appendix (as well as in  \cite{lin_strong_2011, blas_comment_2010}),
we show   clearly  that all these terms can be made either superrenormalizable or strictly renormalizable. }

\end{document}